\documentclass[preprint,12pt,authoryear]{elsarticle}

\usepackage{amssymb}

\usepackage{amsmath}
\usepackage{graphicx}
\usepackage{enumerate}

\usepackage{float}

\usepackage{multibib}
\newcites{sm}{References for Supplementary Materials}
\usepackage{natbib}

\usepackage{url} % not crucial - just used below for the URL

\usepackage[utf8]{inputenc}
\usepackage[english]{babel}

\usepackage{amsmath,amsthm,amssymb}
\usepackage{mathtools}
\usepackage{relsize}
\usepackage{bbm}
\usepackage{bm}

\usepackage{easybmat}
\usepackage{eqparbox}

\usepackage{graphicx}
\usepackage{multirow}
\usepackage{array}
\usepackage{booktabs}
\usepackage{multicol}

\usepackage{subcaption}

\usepackage{pdflscape}

\usepackage{tikz-cd}
\usetikzlibrary{automata,arrows,positioning,calc}

\usepackage[ruled,vlined]{algorithm2e}

\newtheorem{definition}{Definition}
\newtheorem{theorem}{Theorem}
\newtheorem{lemma}{Lemma}
\newtheorem{proposition}{Proposition}
\newtheorem{corollary}{Corollary}
\newtheorem{example}{Example}
\newtheorem{remark}{Remark}

\newtheorem*{lemma*}{Lemma}

\newtheoremstyle{named}{}{}{\itshape}{}{\bfseries}{.}{.5em}{#1\thmnote{ #3}}
\theoremstyle{named}
\newtheorem*{namedlemma}{Lemma}

\DeclareMathOperator*{\argmin}{arg\,min}

\numberwithin{lemma}{section}
\numberwithin{proposition}{section}
\numberwithin{example}{section}

\journal{Computational Statistics and Data Analysis}

\newcommand{\Real}{\mathbb R}

\def\argmin{\mathop{\rm argmin}}

\def\1v{\mathbf 1}
\def\0v{\mathbf 0}
\def\Id{ I}

\begin{document}

\begin{frontmatter}



\title{Integrative decomposition of multi-source data by identifying partially-joint score subspaces}


\author[inst1]{SeoWon Choi}
\ead{chsw1@snu.ac.kr}

\affiliation[inst1]{organization={Department of Statistics, Seoul National University},%Department and Organization
            state={Seoul},
            country={Republic of Korea}}

\author[inst1]{Sungkyu Jung \fnref{thanksto}}
\ead{sungkyu@snu.ac.kr}

\fntext[thanksto]{This work was supported by Samsung Science and Technology Foundation under Project Number
SSTF-BA2002-03.}

\begin{abstract}
Analysis of multi-source dataset, where data on the same objects are collected from multiple sources, is of rising importance in many fields, most notably in multi-omics biology.  A novel framework and algorithms for integrative decomposition of such multi-source data are proposed to identify and sort out common factor scores in terms of whether the scores are relevant to all data sources (fully joint), to some data sources (partially joint), or to a single data source. The key difference between the proposed method and existing approaches is that raw source-wise factor score subspaces are utilized in the identification of the partially-joint block-wise association structure. To identify common score subspaces, which may be partially joint to some of data sources, from noisy observations, the proposed algorithm sequentially computes one-dimensional flag means among source-wise score subspaces, then collects the subspaces that are close to the mean. The proposed decomposition boasts fast computational speed, and is superior in identifying the true partially-joint association structure and recovering the joint loading and score subspaces than competing approaches.
The proposed decomposition is applied to a blood cancer multi-omics data set, containing measurements from three data sources. Our method identifies a latent score, partially joint to the drug panel and methylation profile data sources but not relevant to RNA sequencing profiles, which  helps discovering hidden clusters in the data.
%
% shared signal component that a separate PCA analysis on each single data block could not identify.
%
%
%In numerical studies, our method shows better performance in identifying the underlying partially-joint structure than competing methods in this field.
%
%...
%
%
%One of the major challenges is the partially-joint structure, in which multiple data blocks (all or some of) share a common score matrix, called a partially-joint score. Our goal is to estimate the partially-joint structure by identifying partially-joint scores from factor score subspaces of the given data blocks. We propose a sequential algorithm
%that gathers score subspaces of different data blocks within certain angle threshold
%and identifies partially-shared score, using the concept of angles
%between linear subspaces of different dimensions in $\mathbb{R}^n$.
%
%In numerical studies, our method shows better performance in identifying the underlying partially-joint structure than competing methods in this field. In real data analysis, our method is applied to a blood cancer multi-omics dataset and captures a latent shared signal component that a separate PCA analysis on each single data block could not identify. The proposed method also boasts fast computational speed. Supplementary materials for this article are available online.
\end{abstract}




\end{frontmatter}



\section{Introduction}
\label{sec:intro}

In various fields of science and technology, there is a growing interest in analyzing multi-source data in an integrative way. By the \textit{multi-source data}, we mean data obtained for multiple groups of variables on the same set of subjects. Each group of variables is observed from a common source, and form a data block.
%, $\{ X_k \}_{1 \leq k \leq K}$.
A prominent example of multi-source data is modern multi-omics data that include gene expressions, RNA sequencing, mutations, epigenetic markers or metagenomic materials \citep{Subramanian2020}. The recent development of high-throughput technologies enables us to extract these sources of information comprehensively from a given preparation of cancer/normal tissue samples \citep{Reuter2015, Norris2017}.

One of the main challenges in analyzing multi-omics data is that data blocks come from distinct measurements of different sources.
For example, our motivating data set consists of three data blocks, each from a drug response panel, genome-wide DNA methylation profiles and RNA sequencing profiles \citep{Dietrich2018}; see Section~\ref{sec:realdata} for a detailed description of the data.
% %
%
%
% for example, gene expression or methylation variables in a multi-omics dataset.
Separately analyzing each data block hinders  the assessment of inter-relations among different data blocks. To capture the potentially joint association structures in these multi block data, the \textit{linked component model} has been oftentimes used \citep{Smilde2003,VanDeun2009,Lock2013}.
%It usually guides to a separate analysis on each data block and hinders the assessment of inter-relations among different sources.
%To deal with the situation, a line of works implemented \textit{linked component models} \citep{Smilde2003,VanDeun2009,Lock2013}.

Suppose that there are $K$ data blocks, $X_k \in \mathbb{R}^{p_k \times n}$, for $k = 1, \ldots, K$, observed for $n$ subjects. The subjects in the data are common and matched, i.e., the measurements for the $i$th subject appear in the $i$th column of each data matrix.  Assuming zero mean, each data block is decomposed into
\begin{equation}\label{eq:firstdecomp}
  X_k = Z_k + E_k =  U_k V_k^T + E_k, \quad k = 1, \ldots, K,
\end{equation}
%$$X_k = Z_k + E_k =  U_k V_k^T + E_k = \sum_{j=1}^{r_k} u_{k,i}v_{k,i}^T + E_k, \quad k = 1, \ldots, K,$$
where the low-rank ``signal'' matrix $Z_k$ is factored into a loading matrix $U_k$ and a score matrix $V_k$, which is perturbed by ``noise'' matrix $E_k$. The linked component model further assumes that two or more data blocks can potentially  share a common score component \citep{VanDeun2009}. An extreme example is that all scores are common to all data blocks, that is, $V_1 = \cdots = V_k$, as in \cite{Smilde2003}. The models considered in \cite{Lock2013} and \cite{Feng2018} allow some scores to be common to all data blocks, explaining the joint variation among all data blocks, and some scores to be specific to each data blocks, explaining the individual variations in a single data block.

Following \cite{Gaynanova2019} and \cite{Gao2020}, we in addition allow \emph{partially-joint} scores that are shared across multiple, but not necessarily all, data blocks. An illustrative example of such a model, for $K = 3$ blocks of data, is $ (X_1^T, X_2^T, X_3^T)^T  = UW^T + E$,
\begin{equation}\label{eq:firsteq}
   UW^T  = \begin{pmatrix} U_{(1),1} & U_{(1),2} & 0 \\
U_{(2),1} & U_{(2),2 } & 0 \\
U_{(3),1} & 0 & U_{(3),3} \end{pmatrix}
 \begin{pmatrix} W_{\{1,2,3\}} & W_{\{1,2\}} & W_{\{3\}} \end{pmatrix}^T,
\end{equation}
%
%\begin{align}\label{eq:firsteq}
%  (X_1^T, X_2^T, X_3^T)^T & = UW^T + E,  \\
%   UW^T & = \begin{pmatrix} U_{(1),1} & U_{(1),2} & 0 \\
%U_{(2),1} & U_{(2),2 } & 0 \\
%U_{(3),1} & 0 & U_{(3),3} \end{pmatrix}
% \begin{pmatrix} W_{\{1,2,3\}} & W_{\{1,2\}} & W_{\{3\}} \end{pmatrix}^T
%\end{align}
%
%\begin{align*}
%\begin{pmatrix} X_1 \\ X_2 \\ X_3 \end{pmatrix} = UW^T + E =
%\begin{pmatrix} U_{(1),1} & U_{(1),2} & 0 \\
%U_{(2),1} & U_{(2),2 } & 0 \\
%U_{(3),1} & 0 & U_{(3),3} \end{pmatrix}
% \begin{pmatrix} W_{\{1,2,3\}} & W_{\{1,2\}} & W_{\{3\}} \end{pmatrix}^T + \begin{pmatrix} E_1 \\ E_2 \\ E_3 \end{pmatrix},
%\end{align*}
where $U$ is the loading matrix, $W_{\{1,2,3\}}$ is a matrix of scores that affect all data blocks, and $W_{\{1,2\}}$ is a partially-joint score matrix, affecting only the first two data blocks, but not the third. The scores in $W_{\{3\}} $ are specific to the third data block. Our goal is to delineate such an association structure from a multi-source data.

In this paper, we develop a novel framework and estimation strategy for integrative decomposition of multi-source data, by identifying scores that are fully joint to all data sources, partially joint to some, or specific to a single data source. The framework utilizes the signal score subspace $[V_k]$, a rank $r_k$ subspace of $\Real^n$, spanned by the rows of the signal matrix $Z_k$ in (\ref{eq:firstdecomp}). The rationale for using $[V_k]$ is straightforward: If a common score of rank $r$ is shared by the first two data blocks, then the intersection of $[V_1]$ and $[V_2]$ is non-empty, and is a dimension $r$ subspace.
To capture and summarize the partially-joint block-wise association structure from the score subspaces $[V_k]$, we define the collection of tuples
$$\mathfrak{S} : = \mathfrak{S}(\{[V_k]\}_{k=1,\ldots,K}) = \{(S, r(S)): S \subset \mathcal{K} \},$$
in which a non-empty subset $S$ of $\mathcal{K} = \{1,\ldots, K\}$ denotes a specific pattern of block-wise association, and $r(S)$ denotes the dimension of common scores corresponding to each association pattern $S$. Note that a pattern $S = \{1,2\}$ (for instance) represents that there are common scores of dimension $r(S)$ partially joint to the first and second data sources.
%
% is denoted by , $W_S$ denotes the corresponding common scores (that are either fully-joint, partially-joint or specific to data sources) of rank $r(S)$.
See Section~\ref{sec:statandmath} for a detailed construction of $\mathfrak{S}$, and Section~\ref{sec:theory} for conditions to guarantee the uniqueness of $\mathfrak{S}$.

To identify the partially-joint score structure $\mathfrak{S}$ from noisy observations, we propose to compute the flag mean \citep{Draper2014} of signal score subspaces $[V_k]$'s, and to test whether the mean is indeed close to each signal score subspace. If a mean $w$ is ``close'' to $[V_1]$ and $[V_2]$, but not to other subspaces, then it becomes a  basis of (estimated) $W_{\{1,2\}}$. We use a tunable parameter to determine the closeness between two subspaces. The algorithm, detailed in Section \ref{sec:estim}, is quick in decomposing multi-source data sets, and boasts a superior performance in identifying $\mathfrak{S}$ and in the estimation of the subspaces spanned by the loading matrices and by the common score matrices.

Recently, there has been a growing interest in integrative decomposition of multi-source data \citep{Lock2013, Li2017, Feng2018, Li2018, Gaynanova2019, park2020integrative,Gao2020,lock2022bidimensional}. Among these, \cite{Gaynanova2019} and \cite{Gao2020} also considered modeling partially-joint association structures. However, these authors focused on
the loading matrix $U$, and exploited the source-wise sparse structure of the matrix $U$, as seen in (\ref{eq:firsteq}). On the contrary, we explicitly utilize the signal score matrices $V_k$ in (\ref{eq:firstdecomp}) in identifying partially-joint scores of (\ref{eq:firsteq}). The loading matrix is naturally obtained as a subsequent step in our proposal. Our approach extends the angle-based joint and individual variation identification of \cite{Feng2018}, in which partially joint variations were not considered. In Section~\ref{sec:simul}, we numerically confirm that our proposal finds the true association structure more accurately than the aforementioned decomposition methods.

The rest of article is organized as follows. In Section~\ref{sec:statandmath}, we formally present our integrative decomposition framework and define the partially joint structure $\mathfrak{S}$, followed by the proposed decomposition algorithm and tuning parameter selection procedure in Section~\ref{sec:estim}. In Section~\ref{sec:theory}, conditions to guarantee unique decompositions are discussed. Section~\ref{sec:simul} is devoted to numerical illustrations and comparisons to existing methods. In Section~\ref{sec:realdata},  we demonstrate the use of the proposed method in an analysis of a blood cancer multi-omics data set associated with drug responses, and reveal that the proposal detects a latent signal pattern, partially joint across two, but not all, data sources, which is not easily seen by a separate analysis of individual data blocks. Proofs, technical lemmas, examples and extended numerical results are given in the supplementary material. R codes and data to reproduce the numerical results in this article can be found at \texttt{https://github.com/sungkyujung/psi}.

\section{Statistical framework}
\label{sec:statandmath}

\subsection{Partially-joint structures}

%\begin{comment}

Consider a set of row-centered data matrices $X_k \in \mathbb{R}^{p_k \times n}$ for $k  =  1, \ldots, K$, where the $i$th column of each data matrix corresponds to the same ($i$th) subject. Let $p = \sum_k p_k$. We assume that each $X_k$ is additively decomposed into the rank $r_k$ true signal block $Z_k$ and random errors $E_k$ so that $X_k = Z_k + E_k$.

We consider a model in which the concatenated signal matrix $Z = (Z_1^T, \ldots, Z_K^T)^T$ is decomposed into
\begin{align*}
Z = UW^T,
\end{align*}
where each column of $W$ (corresponding to a latent score) has Euclidean norm $1$, but not necessarily orthonormal.  We require both $U$ and $W$ to have full column rank $r$, and $r < \min(n,p)$ %We assume that the number of columns of $U$ and $W$ is less than $\min(n,p)$.

Our goal is to find a block-wise association structure among $Z_k$'s, in which a latent score (a column of $W$) is linked to multiple, but not necessarily all of, $Z_k$'s. This ``partial-jointness'' of common scores is imposed by assuming
%
 %share a joint signal component. We describe this manner using a word ``partially-joint''.
%
%We give partially-jointness to the model by assuming
block-sparsity on $U$. An example of such model for $K = 3$ blocks of data sets is %For example, when $K = 3$, we can have a model
\begin{equation}\label{eq:mad}
	 \begin{pmatrix} Z_1 \\ Z_2 \\ Z_3 \end{pmatrix} = \begin{pmatrix} U_{(1),1} & U_{(1),2} & 0  \\ U_{(2),1} &  U_{(2),2} & 0 \\ U_{(3),1} & 0 & U_{(3),3}  \end{pmatrix} \begin{pmatrix} W_1 & W_2 & W_3 \end{pmatrix}^T,
\end{equation}
where the blocks $U_{(1),3}$, $U_{(2),3}$ and $U_{(3),2}$ are exactly zero.
In this case, the scores in $W_1$ are shared by all signal blocks $Z_1$, $Z_2$ and $Z_3$ (thus are related to all data sources), whereas $W_2$ is shared only by $Z_1$ and $Z_2$ (that is, partially joint to data sources), and $W_3$ is specific to $Z_3$.

%In this case, $(U_{(1),1}^T U_{(2),1}^T U_{(3),1}^T)^T W_1^T$ is a signal component shared by all $Z_1$, $Z_2$ and $Z_3$ whereas $(U_{(1),2}^T U_{(2),2}^T)  W_2^T$ are shared ``partially'' by $Z_1$ and $Z_2$ and $U_{(3),2} W_3^T$ is the one specific to $Z_3$.
Each block-wise association pattern can be represented by the set of data-source indices, or an \textit{index-set}, a subset of $\{1, \ldots, K\}$, denoted generally by $S$.
For each index-set $S$, the number of columns of $W$ associated with $S$ is called the \textit{rank} of $S$ and is denoted by $r(S)$. %, for the number of the columns of $W$ associated with $S$.
We summarize the block-sparsity via a \textit{partially-joint structure}, which is the collection of pairs $(S,r(S))$ (with $r(S) > 0$). In our example, if the number of columns of $W_1$, $W_2$ and $W_3$ are 2, 1 and 1, respectively, then the partially-joint structure $\mathfrak{S}$ corresponding to (\ref{eq:mad}) is  $((\{1,2,3\},2),(\{1,2\},1), (\{3\},1))$.

As can be seen in the example (\ref{eq:mad}) of the decomposition $Z = UW^T$, for each index-set $S$, the columns of $W$ associated with $S$ are concatenate into an $n \times r(S)$ matrix. These blocks of $W$ are called \textit{partially-joint scores}. The columns of $U$ associated with $S$ are also concatenated into a $p \times r(S)$ matrix, which can be split into block matrices of $p_k \times r(S)$ corresponding to each $Z_k$. We call these block matrices \textit{partially-joint loadings}. If $k \notin S$, then the partially-joint loading matrix corresponding to $Z_k$ is zero. In our example, $W_1$,$W_2$ and $W_3$ are partially-joint scores and $U_{(1),1}$, $U_{(2),1}$, $U_{(3),1}$, $U_{(2),1}$, $U_{(2),2}$ and $U_{(3),3}$  are partially-joint loadings.

For given $Z_k$'s, our aim is to identify partially-joint loadings $U$ and scores $W$. Only with the block-sparsity constraint, there may be multiple candidates of both $U$ and $W$. In the next subsection, we present our strategy of identifying $U$ and $W$ by finding the column space of each $W_i$.

\begin{remark}
The decomposition of $Z_k$'s into the partially-joint loading and score matrices is not unique. If instead of $W_i$ (and its corresponding loading matrix $U_{(k),i}$), one chooses $W_i' = W_iR$, for an orthogonal matrix $R$, then the corresponding loading matrix becomes $U_{(k),i}' = U_{(k),i}R$. Nevertheless, if $U_{(k),i}$ is zero, then $U_{(k),i}R$ is zero for any orthogonal $R$, so the block-sparsity of $U$ is invariant to the choices of basis.
\end{remark}

\subsection{Partially-joint score subspace and index-set ordering}\label{sec:2.2}

We now define the \textit{partially-joint score subspace} $[W_i]$, a linear subspace spanned by the columns of a partially-joint score $W_i$, in a constructive manner. The definition depends mainly on the two concepts: the signal score subspace of signal block $Z_k$ and the ordering of index-sets.

Let the true signal matrix be further decomposed into $Z_k = U_k V_k$, as in the factor analysis, where $V_k \in \mathbb{R}^{n \times r_k}$ is the factor score matrix satisfying $V_k^T V_k = I_{r_k}$. We call the column space of $V_k$, denoted by $[V_k]$, as the \textit{signal score subspace} for $Z_k$. Note that $[V_k]$ is a rank-$r_k$ subspace of $\mathbb{R}^n$, spanned by the columns of $V_k$.

A geometric implication of using the signal score subspaces $[V_k]'s$ in defining partially-joint scores $W_i$'s is as follows:
Take an index set $S = \{1,2\} \subset \{1,\ldots, K\}$ as an example. If $[V_1]$ and $[V_2]$ overlap, then the corresponding signal blocks $Z_1$ and $Z_2$ share a common score, say $W_{\rm cand}$, represented by the subspace $[W_{\rm cand}] := [V_1] \cap [V_2]$.
Such a common score $W_{\rm cand}$ is  at least partially-joint to $Z_1$ and $Z_2$ (and may be related to all or some of $Z_3, \ldots,Z_K$).
%
%For an index set $S \subset \{1,\ldots, K\}$, if $[V_k]$'s $(k \in S)$ overlap, then the corresponding signal blocks $Z_k$, $k \in S$ share a common score, say $W_{\rm cand}$, represented by the subspace $[W_{\rm cand}] := \cap_{k \in S} [V_k]$.  Such a common score $W_{\rm cand}$ is  at least partially-joint to signal blocks $Z_k$ for $k \in S$.
As we have assumed that signal blocks $Z_k$'s share the corresponding partially-joint score, the intersections $\cap_{k \in S} [V_k]$ of signal score subspaces play an important role, as a building block, in identifying partially-joint score subspaces $[W_i]$.

However, one of the challenges of such approach is that index-sets, as subsets of $\{1, \dots, K\}$, have a partially-ordered relation with respect to set-inclusion. As two index-sets can be either nested or intersected, the construction of a partially-joint score subspace corresponding to  an index-set may affect other partially-joint score subspaces. Thus we suggest a plausible way of constructing partially-joint score subspaces by specifying an order among index-sets.

We give an \textit{ordering to index-sets}, sorted by the number of elements in descending order. %Each index-set is allocated with index numbers as $S_i$ for $i = 1, \ldots, 2^K - 1$.
The first index-set is $S_1 = \{1, \ldots, K\}$, the largest set (corresponding the fully-joint score subspace), and each of the next $K$ index-sets $S_2,\ldots,S_{K+1}$ contains exactly $K-1$ indices and so on. (There are $2^{K}-1$ different index-sets.) Note that index-set ordering is not unique for given $K$, since there can be permutations among index-sets of the same size.
Conditions to guarantee invariance of the decomposition with respect to different choices of indexing will be discussed in Section~\ref{sec:theory}.

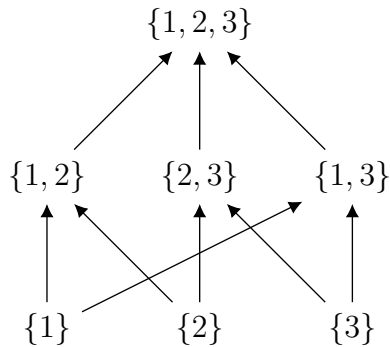
\begin{figure}[bt]
\centering
\resizebox{0.5\columnwidth}{!}{
       \begin{tikzpicture}[> = stealth,     node distance = 2cm]

       %%%% LEVEL 1
\node[scale=0.2][]  (111) {$\{ 1,2,3 \}$};

       %%  LEVEL  2
        \node[scale=0.2][below of=111]  (011) {$\{ 2,3 \}$};
        \node[scale=0.2][right of=011]  (101) {$\{ 1,3 \}$};
       \node[scale=0.2][left of=011]  (110) {$\{ 1,2 \}$};

       %%% LEVEL 3
        \node[scale=0.2] [below of=011] (010) {$\{ 2 \}$};
        \node[scale=0.2][below of=101]  (001) {$\{ 3 \}$};
        \node[scale=0.2] [below of=110] (100) {$\{ 1 \}$};

%% LEVEL 1 to LEVEL 2
		\draw[->,line width=0.1pt, arrows = {-Latex[width=1pt, length=1pt]}] (011)  edge node {} (111);
		\path[->,line width=0.1pt, arrows = {-Latex[width=1pt, length=1pt]}] (101)  edge node {} (111);
		\path[->,line width=0.1pt, arrows = {-Latex[width=1pt, length=1pt]}] (110)  edge node {} (111);

%% LEVEL 2 to LEVEL 3
		\path[->,line width=0.1pt, arrows = {-Latex[width=1pt, length=1pt]}] (010)  edge node {} (011);
		\path[->,line width=0.1pt, arrows = {-Latex[width=1pt, length=1pt]}] (001)  edge node {} (011);
		\path[->,line width=0.1pt, arrows = {-Latex[width=1pt, length=1pt]}] (100)  edge node {} (101);
		\path[->,line width=0.1pt, arrows = {-Latex[width=1pt, length=1pt]}] (001)  edge node {} (101);
		\path[->,line width=0.1pt, arrows = {-Latex[width=1pt, length=1pt]}] (100)  edge node {} (110);
        \path[->,line width=0.1pt, arrows = {-Latex[width=1pt, length=1pt]}] (010)  edge node {} (110);
      \end{tikzpicture}
}
\caption{An example of index-set ordering is depicted as the indexed partially-ordered set. Arrow $\rightarrow$ stands for $\subset$, e.g. $\{1\} \rightarrow \{1,2\}$ means $\{1\} \subset \{1,2\}$.}\label{fig:ordering}
\end{figure}

\begin{example}\label{exam:examone}
 For $K = 3$, $S_1 = \{1,2,3\}$ as appeared in the top row of Fig.~\ref{fig:ordering}, and the next three sets ($S_2, S_3, S_4$) are given by the next row of the figure. That is, $S_2$, $S_3$, and $S_4$ are $\{ 1, 2\}$, $\{ 2, 3\}$, $\{ 1, 3\}$, respectively. Likewise, $S_5 = \{ 1\}$, $S_6 = \{ 2 \}$, $S_7 = \{ 3 \}$.  Elements in $S_i$ stands for the indices of data blocks that potentially possess a common score.
 Such an ordering is not unique, and one may set, e.g., $S_6 = \{ 3 \}$, $S_7 = \{ 2 \}$.
\end{example}

Using the signal score subspaces $[V_k]$'s and a pre-specified index-set ordering, we now define partially-joint score subspaces.

\begin{definition}\label{def:score}
Suppose that data matrices $X_k = Z_k + E_k \in \mathbb{R}^{p_k \times n}$ for $k = 1, \ldots, K$ with true signal block $Z_k$ are given, in which the data matrices are matched, i.e., the $j$th row of each $X_k$ represents the measurements for the $j$th individual. With an index-set ordering $S_1, \ldots, S_{2^K-1}$, a set of corresponding partially-joint score subspace $[W_i]$'s for $i = 1, \ldots, 2^K-1$ are defined sequentially:
\begin{align*}
[W_1] &:= \cap_{k \in S_1 } [V_{k}], \\
[W_2] &:= P_1^\perp \left( \cap_{k \in S_i} [V_{k}] \right), \\
&\vdots \\
[W_i] &:= ( \underset{ \substack{ \{ j : j < i, \\ S_j \cap S_i \neq \phi \} } }{\bigcirc}  P_j^\perp ) \left( \cap_{k \in S_i} [V_{k}] \right).
\end{align*}
Here, $P_j^{\perp}$ is the projection transformation of $\mathbb{R}^n$ onto the orthogonal complement of $[W_j]$. The notation  $\bigcirc_{j \in J} P_j^\perp ([A])$, for a set $J \subset \{ 1, \ldots, 2^K - 1 \}$ and a subspace $[A]$, stands for the repeated applications of $P_j^{\perp}$ on $[A]$, where $P_j^\perp$s are applied one by one by the increasing order of $j \in J$.
\end{definition}
 Note that depending on $\{ [V_k] \}$, a $[W_i]$ may be $\{ 0 \}$. (As a convection, ${\rm rank}(\{ 0 \}) = 0$.)
The block-wise association structure $\mathfrak{S}$ for the matched data $X_k$'s is then the collection of $(S_i, r(S_i))$ for which $r(S_i) = {\rm rank}([W_i]) > 0$.

\begin{example} \label{exam:examtwo}
Suppose $K = 3$ and the index-set ordering is given as in Example \ref{exam:examone}, with $r(S_1) = r(S_2) = r(S_4) = r(S_7) =  1$, and $r(S_i) = 0$ otherwise. Partially-joint score subspaces are obtained as $W_1 = \cap_{k \in S_1} [V_k]$, $W_2 = P_1^\perp(\cap_{k \in S_2} [V_l])$, $W_4 = P_2^\perp \circ P_1^\perp(\cap_{k \in S_4} [V_k])$ and $W_7 = P_4^\perp \circ P_1^\perp(\cap_{k \in S_7} [V_k])$.
\end{example}

 Partially-joint score subspaces have the following basic properties.
 \begin{enumerate}
 \item[(1)] The partially-joint score subspaces involved with each $Z_k$ can be used to restore the signal score subspace. That is,
\begin{equation}\label{eq:subspacesum_orthogonal}
  [V_k] = +_{i \in \{ k \in S_i\} } [W_i] ,
\end{equation}
as shown in Lemma~A.3 in the supplementary material. Here the notation `+' means the sum of subspaces.
 \item[(2)] For any $i \neq j$, $[W_i]$ and $[W_j]$ do not overlap, i.e.,  $[W_i] \cap [W_j] = \{0\}$, but they are not necessarily orthogonal. However, if both $[W_i]$ and $[W_j]$ are related to a common data source, then they are orthogonal, as shown in the following lemma.
 \end{enumerate}
 \begin{lemma} \label{lem:lemone}
	For $i, j \in \{1,\ldots,2^{K}-1\}$ and $S_i \cap S_j \neq \phi$, $[W_i] \perp [W_j]$.
\end{lemma}
 Lemma \ref{lem:lemone} ensures that all partially-joint score subspaces relevant to the $k$th data block, $\{[W_{i}] : k \in S_i, i = 1,\ldots, 2^{K}-1\}$, are orthogonal to each other.  Thus,
the subspaces in the right hand side of (\ref{eq:subspacesum_orthogonal}) are  indeed orthogonal to each other.

\subsection{Partially-joint score and loading}

Given $\mathfrak{S}$,
choose an orthogonal basis $W_i \in \mathbb{R}^{n \times r(S_i)}$ of $[W_i]$, which becomes a partially joint score. Let $W_{(k)}$ be the column-wise concatenation of $W_i$'s for $i \in \{ i : k \in S_i \}$. That is, $W_{(k)}$ collects the partially-joint scores corresponding to the $k$th data source.  By Lemma~\ref{lem:lemone}, the matrix $W_{(k)}$ is an orthogonal matrix. The partially-joint loadings corresponding to $Z_k$ are then given by
\begin{align*}
U_{(k)} = Z_k W_{(k)},
\end{align*}
where $U_{(k)}$ is the column-wise concatenation of $U_i$'s for $i \in \{ i : k \in S_i \}$. Indeed we can easily check that $U_{(k)} W^T_{(k)} = Z_k W_{(k)}  W_{(k)}^T = Z_k$, since $W_{(k)} W_{(k)}^T$ is the projection matrix onto $[V_k]$ by the property (1)
%Equation (\ref{eq:subspacesum_orthogonal})
above.
The partially-joint loading $U_{(k),i}$ for $i \in \{ i : k \in S_i \}$ is the $(k,i)$th block of  $U$. For  $i \not\in \{ i : k \in S_i \}$, $U_{(k),i} = 0$.

\section{Estimation}
\label{sec:estim}

In practice, the signal $Z_k$ and error $E_k$ of row-centered data matrices $X_k \in \mathbb{R}^{p_k \times n}$ for $k = 1, \ldots, K$ are unknown. We assume that the ranks $r_k$ of the signal matrices are pre-determined and extract the signal matrix using a low-rank approximation of $X_k$, and write the rank $r_k$ approximation of $X_k$ by $\widehat{Z}_k$. The basis of the (empirical) signal score subspace $\widehat{V}_k$ is given either by the right singular vectors of $\widehat{Z}_k$ from or by any off-the-shelf factor model estimates. We use the singular value decomposition (SVD) for our numerical examples.

An overview of the estimation algorithm is as follows: (1) on the outer loop, we iterate through all index-sets $S_i$ on a given ordering, and (2) on the inner loop, we obtain the \textit{partially-joint score subspace estimate} $[\widehat{W}_i]$ from $\{ [\widehat{V}_k] \}_{k \in S_i}$, identifying the basis of $[\widehat{W}_i]$ one-by-one.

A major challenge in doing so is that we only have the sample versions of signal blocks, $\widehat{Z}_k$, and thus the available sample signal score subspaces $[\widehat{V}_k]$ is a perturbed version of the true signal score subspaces $[V_k]$. That is, even if $[V_k]$'s overlap (i.e., $\cap_{k \in S_i} [V_k] \neq \{ 0 \}$) the sample version may not overlap: $\cap_{k \in S_i} [\widehat{V}_k] = \{ 0 \}$. Thus, there is a need to give a slack on identifying the ``intersection'' of $[V_k]$'s, accounting for random perturbations  in $[\widehat{V}_k]$. We propose to use principal angles between subspaces for such identification, further developed in Section~\ref{sec:3.1}.

In Section~\ref{sec:3.1}, we propose a procedure to estimate a basis matrix $\widehat{W}_i$, which gives the partially-joint score subspace estimate $[\widehat{W}_i]$ at each $i$th step. The rank of the corresponding $S_i$ is set as $\hat{r}(S_i) = \textrm{rank}(\widehat{W}_i)$.
Finishing the iteration over all $S_i$'s, an estimated partially-joint structure $\widehat{\mathfrak{S}} = \{ (S_i, \hat{r}(S_i) \}$ and a set of corresponding $[\widehat{W}_i]$'s are obtained. In Section~\ref{sec:3.2}, we then discuss the estimation of corresponding partially-joint loading matrices from $\widehat{\mathfrak{S}}$ and $[\widehat{W}_i]$'s. In %Section~\ref{sec:3.3},  a strategy for thresholding at the right principal angle is proposed.

\subsection{Partially-joint Score Subspace Estimation}\label{sec:3.1}

Our strategy for estimating partially-joint score subspace $[\widehat{W}_i]$, for an index-set $S_i$, is by collecting one-dimensional bases, that lie ``close'' to each and every score subspaces in $\{[\widehat{V}_k]\}_{k \in S_i}$.

For the notion of closeness, the principal angle between subspaces is used, and we will determine that two subspaces are close if the principal angle is smaller than a threshold $\lambda \in [0, \pi/2)$. For now, $\lambda$ is treated as a pre-determined tuning parameter, and a data-driven choice of $\lambda$ will be discussed in Section \ref{subsec:param}.

For a given $\lambda \in [0, \pi/2)$, we  estimate $[\widehat{W}_i]$ sequentially for $i = 1, \ldots, 2^K -1$, using the estimation algorithm presented below.
 At the $i$th stage, we begin with $\mathcal{W}_i= \emptyset$, to which the identified one-dimensional bases of the partially-joint score subspace are added. Algorithm \ref{alg:algone} summarizes the proposed partially-joint structure identification procedure, using steps (a)---(c). \\

\begin{algorithm}[pt]
	\SetAlgoLined
	\SetKwInOut{Input}{input}
	\Input{$\widehat{V}_1$, \ldots, $\widehat{V}_K$, $S_1,\ldots,S_{2^K-1}$ ,  $\lambda$}
	\For{$i = 1, 2, \ldots, 2^K - 1$} {
		Set $\mathcal{W}_i = \phi$\;
		\While{${\rm dim}([\widehat{V}_{k}]) > 0$ {\rm for all} $k \in S_i$} {
			(a) Compute the mean direction $\hat{w}$  of $\{[\widehat{V}_k]\}_{k \in S_i}$,\textit{see} (\ref{DefPD})\;
			\uIf{the condition (\ref{rule0}) is satisfied}{
				(b) Let $\mathcal{W}_i \leftarrow \mathcal{W}_i \cup \{\hat{w}\}$\;
				(c) Update $\widehat{V}_k \leftarrow \widehat{V}_{k, trunc}$ for each $k \in S_i$
			} \Else{
				break\;
			}
		}
        Let $\hat{r}(S_i) = |\mathcal{W}_i|$,
		write $\widehat{W}_i$ for the $n \times \hat{r}(S_i)$ matrix consisting of elements in $\mathcal{W}_i$ and record $(\widehat{S_i}, \hat{r}(S_i), \widehat{W}_i)$ in $\widehat{\mathfrak{S}}$\;
	}
	\KwResult{ $\widehat{\mathfrak{S}}$}
	\caption{Partially-joint Structure Identification}
	\label{alg:algone}
\end{algorithm}

\textit{Step (a)}: %As a candidate that may be included in $\widehat{W}_i$, find a \textit{mean direction} that lies closest to all score subspaces $\{[\widehat{V}_k]\}_{k \in S_i}$, as follows.
%
%As a candidate for a basis vector of the partially joint score matrix,
As a candidate that may be included in $\widehat{W}_i$,
compute the mean direction $\hat{w}$ among $\{ [\widehat{V}_k] \}_{k \in S_i}$, provided that $[\widehat{V}_k] \neq \{0\}$ for all $k \in S_i$.
The mean direction minimizes the sum of the squares of the subspace distances between a candidate $w$ and subspaces $[\widehat{V}_k]$ for $k \in S_i$, and is
	\begin{align} \label{DefPD}
	\hat{w} = \argmin_{ w^T w = 1 } \sum_{k \in S_i} d \left( [w], [\widehat{V}_k] \right)^2,
	\end{align}
	where $d([w], [B]) = 1 / \sqrt{2} \cdot \Vert w w^T - B B^T \Vert_F$ is the Frobenius-norm distance between subspaces $[w]$ and $[B]$  \citep{Ye2016}. We chose the Frobenius norm, since for any choice of the basis $\widehat{V}_k$ for $[\widehat{V}_k]$,
\begin{align*}
 \sum_{k \in S_i} d \left( [w], [\widehat{V}_i] \right)^2 = |S_i| - w^T \left( \sum_{k \in S_i} \widehat{V}_k \widehat{V}_k^T  \right) w = |S_i| -  w^T (H H^T) w,
\end{align*}
where $H$ is the matrix given by the column-wise binding of $\widehat{V}_k$'s, and $\hat{w}$ is
the first left singular vector of $H$. We mention in passing that $[\hat{w}]$ is the one-dimensional flag mean of subspaces $\{ [\widehat{V}_k] \}_{k \in S_i}$ \citep{Draper2014}. \\

\textit{ Step (b)}: Check whether all of the signal score subspaces $[\widehat{V}_k]$ are not too dispersed from $\hat{w}$ in (\ref{DefPD}).
For this, given the prespecified $\lambda$, we check whether the \emph{principal angle} between the mean direction and each of $[\widehat{V}_k]$ is at most $\lambda$, i.e.,
 %   \begin{equation}\label{rule0}
 %   d([\hat{w}],[\widehat{V}_k]) < \sin(\lambda), \quad \mbox{for all}\  k \in S_i.
 %   \end{equation}
    \begin{equation}\label{rule0}
    \theta([\hat{w}],[\widehat{V}_k]) <  \lambda, \quad \mbox{for all}\  k \in S_i.
    \end{equation}
Here, the principal angle $\theta([w] ,[B]) := \arcsin ( d \left( [w], [B] \right) ) \in [0, \pi/2]$ is the acute angle formed by the vector $w$ and the subspace $[B]$ \citep{Bjorck1973}. If the condition (\ref{rule0}) is not satisfied, then skip the following and move to the next stage for $S_{i+1}$. If  (\ref{rule0})  is satisfied, then
$\mathcal{W}_i$ is updated to $\mathcal{W}_i \leftarrow \mathcal{W}_i \cup \{\hat{w}\}$. \\

\textit{Step (c)}: Deflate each $[\widehat{V}_k]$ for $k \in S_i$, so that the next mean direction is orthogonal to previous ones.
In particular, we ``peel'' the basis $\hat{w}$, that has been added to $\mathcal{W}_i$, from each of $[\widehat{V}_k]$, $k \in S_i$. Since $\hat{w}$ is not exactly in $[\widehat{V}_k]$, the one-dimensional subspace closest to $\hat{w}$ is removed. Specifically,
 let $[\widehat{V}_{k,trunc}]$ be the orthogonal complement of $P_{\widehat{V}_k} \hat{w}$ within   $[\widehat{V}_k]$. (Note that  ${\rm dim}([\widehat{V}_{k,trunc}]) = {\rm dim}([\widehat{V}_{k}]) -1$.)
 We update $[\widehat{V}_k]$ by $[\widehat{V}_k] \leftarrow [\widehat{V}_{k,trunc}]$,
 %Write $[\widehat{V}_k]$ for $[\widehat{V}_{k,trunc}]$,
 for each $k \in S_i$. If any $[\widehat{V}_k]$ for $k \in S_i$ becomes $\{0\}$, then move to the next stage for $S_{i+1}$. Otherwise, move back to step (a). \\

We give an illustrative example for the three steps.

\begin{figure}[tb]
	\begin{center}
		\includegraphics[trim={0 2.5cm 0 2.5cm},clip,width=3.5in]{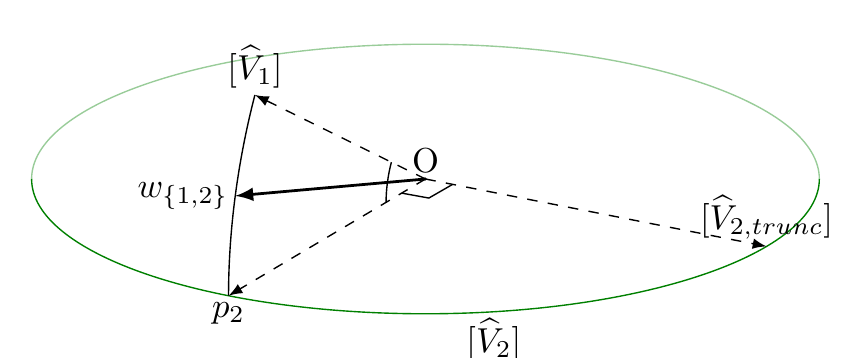}
	\end{center}
	\caption{A figurative description for computing partially-joint score subspace $[\widehat{W}_i]$, for $S_i = \{1,2\}$. The two-dimensional subspace $[\widehat{V}_2]$ is depicted as a disk. $w_{\{1,2\}}$ is the mean direction and $p_2$ stands for the projection of $w_{\{1,2\}}$ onto $[\widehat{V}_2]$. \label{fig:first}}
\end{figure}

\begin{example}
Suppose $n = 3$ and $K = 3$. Assume that we are at the stage $S_i = \{1,2\}$ and $[\widehat{V}_1]$ has been deflated (in the previous stages) to be of dimension 1, while $[\widehat{V}_2]$ is of dimension 2, as in Fig.~\ref{fig:first}. There, $[\widehat{V}_1]$ is generated by $(\cos \pi/6 ,0,\sin \pi/6 )^T$, and $[\widehat{V}_2]$ by $(1,0,0)^T$ and $(0,1,0)^T$.
In Step (a), the mean direction is $\hat{w} = (\cos  \pi/ 12 ,0,\sin  \pi / 12 )^T$. In Step (b), for $k = 1,2$, $\theta([\hat{w}], [\widehat{V}_k]) = \pi / 12$, (\ref{rule0}) is satisfied for any tuning parameter  $\lambda > \pi/12$. Thus we say that $[\widehat{V}_1]$ and $[\widehat{V}_2]$ share a partially-joint score subspace, and  $\widehat{w}$ is included as a basis of $[W_i]$.
In step (c), $[\widehat{V}_2]$ is updated to $[\widehat{V}_2] \leftarrow [\widehat{V}_{2,trunc}] = \textrm{span}((0,1,0)^T)$, and $[\widehat{V}_1]$ becomes $\{0\}$. Since there is no more scores left to exploit, we move onto the next stage for $S_{i+1} = \{2,3\}$.
\end{example}

For a singleton set $S_i  = \{ k \}$,
the mean direction $\hat{w}$ is any unit vector in (the deflated) $[\widehat{V}_k]$, and the condition (\ref{rule0}) is always satisfied. Thus, for this case, $\widehat{W}_i = \widehat{V}_k$, in place of steps (a)---(c) above.

It can be seen that the estimated partially-joint score matrices $\widehat{W}_i$ consist of orthogonal columns, and $\widehat{W}_i^T\widehat{W}_i = \Id_{\hat{r}(S_i)}$, because the deflated subspaces in Step (c) above are orthogonal to $\hat{w}$, and the next mean direction lied in the union of the deflated subspaces.

\subsection{Partially-joint Loading Matrix Estimation} \label{subsec:estimU}\label{sec:3.2}

Given the partially-joint structure estimate $\widehat{\mathfrak{S}} = \{ (S_i, \hat{r}(S_i)) : i \in \mathcal{I}_K \}$ and the corresponding partially-joint score subspace estimates $[\widehat{W}_i]$'s, we obtain the estimated partially-joint loading matrix $\widehat{U}$.

Let $\widehat{Z} = (\widehat{Z}_1^T, \ldots, \widehat{Z}_K^T)^T \in \mathbb{R}^{p \times n}$, where $p = \sum_{k=1}^K p_k$.
Denote the column-wise concatenation of $\widehat{W}_i$ as $\widehat{W} \in \mathbb{R}^{n \times \hat{r} }$, where $\hat{r} = \sum_{i \in \mathcal{I}_K} \hat{r}(S_i)$.

We estimate the partially-joint loading matrix from the optimization problem
\begin{align} \label{rule4}
\widehat{U} = \argmin_{U \in \mathbb{R}^{ p \times \hat{r} }} \Vert \widehat{Z} - U \cdot \widehat{W}^T   \Vert_F^2.
\end{align}
with a constraint that $U$ in (\ref{rule4}) has the block-wise sparsity structure that corresponds to the estimated partially-joint structure $\widehat{\mathfrak{S}}$.
For example, given $K = 2$ and $\widehat{\mathfrak{S}} = \{ (\{1,2\} , 2), (\{ 1 \} , 1) \}$, $U$ has a block sparsity structure
\begin{align*}\label{exam:examfour}
\begin{BMAT}(r)[3pt,0pt,2.0cm]{c}{cc}
p_1 \\
p_2
\end{BMAT}
\begin{BMAT}(r)[-2pt,0pt,2.0cm]{c}{cc}
\left\lbrace \vphantom{\rule{1mm}{14pt}} \right. \\
\left\lbrace \vphantom{\rule{1mm}{14pt}} \right.
\end{BMAT}
\left[
\begin{BMAT}(e)[2pt,2.0cm,2.0cm]{c.c}{c.c}
U_{(1),1} & U_{(1),2}  \\
U_{(2),1} & 0
\end{BMAT}
\right]
\end{align*}
with the number of columns of $(U_{(1),1}^T, U_{(2),1}^T)^T$ and $U_{(1),2}$ are two and one, respectively.

With the block-wise sparse constraint imposed, the objective function (\ref{rule4}) for $U$ can be written separately for each data block, i.e.,
\begin{equation}\label{eq:constrainedLoading}
  \Vert \widehat{Z} - U \cdot \widehat{W}^T  \Vert_F^2
  = \sum_{k = 1}^K \Vert \widehat{Z}_k -  U_{(k)} \widehat{W}_{(k)}^T \Vert_F^2.
\end{equation}
%\begin{align}\label{eq:constrainedLoading}
%  \Vert \widehat{Z} - U \cdot \widehat{W}^T  \Vert_F^2
%  = \sum_{k = 1}^K \Vert \widehat{Z}_k -  U_{(k)} \widehat{W}_{(k)}^T \Vert_F^2
%\end{align}
Here, $U_{(k)}$ and $\widehat{W}_{(k)}$ are the column-wise concatenation of each $U_{(k),i}$'s and $\widehat{W}_i$'s, respectively, for $i \in  \{ i : k \in S_i\  \mbox{and} \ \widehat{r}(S_i) > 0\}$. The minimizer of (\ref{eq:constrainedLoading}) is
$$
\widehat{U}_{(k)} = \widehat{Z}_k \widehat{W}_{(k)} (\widehat{W}_{(k)}^T \widehat{W}_{(k)})^{-1},
$$
and $\widehat{U}_{(k),i}$ for $i \in \mathcal{J}_{(k)}$ are obtained by disjoining $\widehat{U}_{(k)}$. By the block-wise sparse structure, set $\widehat{U}_{(k),i} = 0$ if $k \notin S_i$. %The estimated partially-joint loading matrix consists of $\widehat{U}_{(k)i}$, and is denoted by $\widehat{U} \in \Real^{p \times \hat{r}}$.

\subsection{Tuning Parameter Selection} \label{subsec:param}\label{sec:3.3}
The partially-joint structure identification, proposed in Section 3.1, depends heavily on the tuning parameter  $\lambda \in [0, \pi/2)$. If $\lambda$ is too small, then all scores are identified as individual scores, specific to each data blocks. If $\lambda$ is too large, then individual and partially-joint scores may be falsely identified as fully-joint scores.

We use data splitting to select the value of tuning parameter $\lambda \in [0, \pi/2)$. For a single instance of data splitting, randomly split $n$ samples of $X = [X_{1}^T, \ldots, X_{K}^T]^T$ into two groups of equal proportions, the training set $X_{tr} = [X_{tr,1}^T, \ldots, X_{tr,K}^T]^T$ and the test set $X_{test} = [X_{test,1}^T, \ldots, X_{test,K}^T]^T$.

Given the signal rank $r_k$ of each $X_k$,
we then extract the training signal matrices $\widehat{Z}_{tr,k}$ for $k = 1, \ldots, K$ using the rank $r_k$ approximation of $X_{tr,k}$.
For each $\lambda$ on the tuning parameter grid, we identify the partially-joint structure from $\widehat{Z}_{tr,k}$'s, and
obtain the partially-joint score $\widehat{W}_{tr, \lambda}$ and the partially-joint loading matrix $\widehat{U}_{tr, \lambda}$, as discussed in Sections 3.1 and 3.2.

To assess the degrees to which the estimates are generalized to the test set, we first evaluate the score matrix for the test set, given by the loading matrix estimates $\widehat{U}_{tr, \lambda}$ from the training data. The test score matrix $\widehat{W}_{test, \lambda}$ is defined as the minimizer $\widehat{W}_{test,\lambda} \in \mathbb{R}^{n_{test} \times \hat{r}}$ of $$\Vert X_{test} - \widehat{U}_{tr, \lambda} W^T \Vert_F^2$$ subject to $W^T W = I_{\hat{r}}$.
%W_{tr,\lambda}^T W_{tr,\lambda}$.
 The test score matrix is then
$\widehat{W}_{test,\lambda} = P_\lambda Q_\lambda^T$, where $P_\lambda$ and $Q_\lambda$ are left and right singular vector matrices of $X_{test}^T \cdot \widehat{U}_{tr, \lambda} = P_\lambda \Sigma_{\lambda} Q_\lambda^T$ using the Eckart-Young theorem \citep{Eckart1936}.
We remark that the constraint $W^TW = I_{\hat{r}}$ assumes orthogonality among the scores, which may not be satisfied. Other options include imposing $W^TW = \widehat{W}_{tr, \lambda}^T\widehat{W}_{tr, \lambda}$, thereby requiring the correlations in the test scores to be the same as those in the training scores. Due to high variability of $\widehat{W}_{tr, \lambda}^T\widehat{W}_{tr, \lambda}$ over random splitting, this approach generally leads to unstable choice of $\lambda$. 

Let $\tilde{\lambda}$ be the value of $\lambda$ for which the minimum of an empirical risk is attained. The empirical risk, defined for $\lambda \in [0, \pi/2]$, is
\begin{align}\label{eq:risk}
\textrm{Risk} (\lambda) = \sum_{k = 1}^K \frac{ \Vert X_{test,k} - \widehat{U}_{tr, \lambda,(k)} \widehat{W}_{test,\lambda}^T \Vert_F^2 }{\Vert X_{test,k}  \Vert_F^2},
\end{align}
where $\widehat{U}_{tr, \lambda, (k)}$ is the $k$th row block of $\widehat{U}_{tr, \lambda}$. A similar form was used in \citet{Gaynanova2019}.
The corresponding partially-joint structure is
$\widehat{\mathfrak{S}}_{tr}(\tilde{\lambda})$. 

Note that the tuned parameter $\tilde{\lambda}$ is optimal in the sense that it minimizes (\ref{eq:risk}). However, as $\tilde{\lambda}$ is tuned using only a half of the sample (with greater variability of $\hat{w}$ and $[\widehat{V}_i]$), directly using $\tilde{\lambda}$ as the angle threshold (\ref{rule0}) for the whole data (with doubled sample size) generally results in identifying too many common scores. On the other hand, the estimated structure
$\widehat{\mathfrak{S}}_{tr}(\tilde{\lambda})$ corresponding to the tuned parameter $\tilde{\lambda}$ does not heavily depend on the sample size. Therefore, we use the chosen structure $\widehat{\mathfrak{S}}_{tr}(\tilde{\lambda})$, rather than the parameter $\tilde{\lambda}$, in determining the angle threshold used in (\ref{rule0}) for the whole data. 
%Finally, we generalize $\widehat{\mathfrak{S}}_{tr}(\tilde{\lambda})$ to the whole data. 
Let $\widehat{Z}_k$ be the rank $r_k$ approximation of the whole data $X_k$. Again on the grid of $\lambda$'s, we obtain the partially-joint structure $\mathfrak{S}(\lambda)$ from $\widehat{Z}_k$'s. Then we choose the best value $\widehat{\lambda}$ that minimize $d (\widehat{\mathfrak{S}}_{tr}(\tilde{\lambda}), \mathfrak{S}(\lambda))$, where $d$ is a measure of dissimilarity between two structures defined below.

\subsubsection{Measure of Dissimilarity}

In generalizing $\widehat{\mathfrak{S}}_{tr}(\tilde{\lambda})$ to the whole data, we compute dissimilarity measure between $\widehat{\mathfrak{S}}_{tr}(\tilde{\lambda})$ and $ \mathfrak{S}(\lambda)$ for each candidate $\lambda$.
For this, we represent each element $(S,r(S)) \in \mathfrak{S}$ by $r(S)$ copies of a binary vector $a_S \{0,1\}^K$ where the $i$th element of $a_S$ is 1 if $i \in S$, 0 otherwise. For example, for $K = 2$, $(S,r(S)) = (\{1,2\},2 )$ is represented by $\{(1,1), (1,1)\}$. Collecting all such binary vectors for each element of $\mathfrak{S}$, as an instance  $\mathfrak{S} = ((\{1,2\},2 ),(\{1\},1 ))$ is represented by the multi-set $\{ (1,1), (1,1), (1,0)\}$.
%Here we define a measure of dissimilarity using $K$-row binary structure matrices, for example, when $K = 2$, $((\{1,2\},2 ),(\{1\},1 ))$ corresponds to
%\begin{align*}
%\begin{pmatrix} 1 & 1 & 1 \\ 1& 1 & 0 \end{pmatrix}.
%\end{align*}
%Our strategy is that we consider the binary structure matrix as a multi-set of binary column vectors (of size $K$).

We devise a measure that swiftly captures dissimilarity between two multi-sets, under a situation that we need to compute dissimilarity as many time as the number of $\lambda$'s on the grid for each instance of splitting.
Let $d_H(a,b)$ be the number of elements between column vectors $a,b \in \{0,1\}^K$, i.e. Hamming distance, which we will use as a ground distance. The dissimilarity measure $d(\mathfrak{S},\mathfrak{S}')$ between two partially-joint structure $\mathfrak{S}$, $\mathfrak{S}'$ is defined as follows
\begin{align*}
d(\mathfrak{S},\mathfrak{S'}) = \sum_{a \in A \setminus B}d_H^2(a,B \setminus A) + \sum_{b \in B \setminus A} d_H^2(b, A \setminus B),
\end{align*}
where $A$ and $B$ are the multi-sets corresponding to $\mathfrak{S}$ and $\mathfrak{S}'$, respectively.
% The set minus operation, denoted $\setminus$, is defined on a class of binary matrices of $K$ rows, regarded as a multi-set of binary columns.
See Section B.1 of the supplementary material for an exemplary description. Such a distance between multi-sets has been appeared in the literature. In fact,  $d$ is a special version of the matching distance between multi-sets defined in \cite{Bolt2022}.

\begin{remark}
It should be noted that $d$ is not a metric, but a semi-metric. The conditions (1) $d(\mathfrak{A},\mathfrak{B})$ if and only if $\mathfrak{A} = \mathfrak{B}$ and (2) $d(\mathfrak{A},\mathfrak{B}) = d(\mathfrak{B},\mathfrak{A})$ hold, but (3) the triangular inequality does not hold in general.
\end{remark}

\section{Theory}
\label{sec:theory}

Given an ordering of index-sets, defined in Section~\ref{sec:2.2} and the signal blocks $Z = \{ Z_k : k  = 1,\ldots, K \}$, the partially-joint structure $\mathfrak{S}(Z) = \{(S_i, r(S_i), [W_i]) : i = 1,\ldots, 2^{K}-1\}$ is uniquely determined. (We abuse the notation $\mathfrak{S}(Z)$ to include the partially-joint score subspaces $[W_i]$'s, for this section.)
Unfortunately, for different orderings given to index-sets the ranks $r(S_i)$ and the partially-joint score subspaces $[W_i]$ may be different. In this section, we introduce conditions on relations among $[V_k]$'s for $\mathfrak{S}(Z)$ to be uniquely determined regardless of the choice of the orders of index-sets provided that they obey the partially-ordered relation with respect to set-inclusion. % $(\mathbbm{2}^{\mathcal{K}} \setminus \{\phi \}, \subset)$.

Let $\mathcal{I}_K = \{1,\ldots, 2^{K}-1\}$.
For each $l = 1, \ldots, K$, let $\mathcal{J}_l = \{ i \in \mathcal{I}_K   : |S_i| = l \}$ be the set of all indices with size $l$. For example, $\mathcal{J}_l$ corresponds to the $l$th row of Figure~\ref{fig:ordering}.
 For $l = 1, \ldots, K-1$, we set $[I_l] = +_{i \in \{ i : |S_i| > l \}} ( \cap_{k \in S_i} [V_{k}] ) = +_{i \in \{ i : |S_i| > l \} } [W_i]$, and the projection onto $[I_l]^\perp$ in $\mathbb{R}^n$ is denoted by $P_{I_l}^\perp$. Note that when evaluating the rank for $[W_i]$ for $i \in \mathcal{J}_l$, the definition of $[W_i]$ in Definition \ref{def:score} utilizes the deflated score subspaces orthogonal to $[I_l]$, using the given order of $S_i$'s among $i \in \mathcal{J}_l$.

Let $\mathcal{K} = \{1,\ldots, K\}$.
 We say $\{ [V_k] \}_{k \in \mathcal{K}}$ to be \textit{relatively independent} if, for every $l = 1, \ldots, K - 1$ and $i \in \mathcal{J}_l$, $P_{I_l}^\perp (\cap_{k \in S_i} [V_k])$ is linearly independent to $[C_{l,-i}] := +_{ j \in \mathcal{J}_l \setminus \{ i \} } (P_{I_l}^\perp ( \cap_{k \in S_i} [V_{k}]))$.
 In words, $\{ [V_k] \}_{k \in \mathcal{K} }$ is relatively independent, if for each and every layer $\mathcal{J}_l$, each deflated subspace is linearly independent to $[C_{l,-i}]$, which is the sum of the other deflated subspaces in the layer $\mathcal{J}_l$.

 If $P_{I_l}^\perp (\cap_{k \in S_i} [V_k])$ is orthogonal to $[C_{l,-i}]$ for every $l = 1, \ldots, K - 1$ and $i \in \mathcal{J}_l$, then $\{ [V_k] \}_{k \in \mathcal{K}}$ is said to be \textit{relatively orthogonal}. We immediately check that relative orthogonality implies relative independence.

\begin{theorem}\label{thm:thmone}
	Given matched data matrices $X_k = Z_k + E_k \in \mathbb{R}^{p_k \times n}$ for $k = 1, \ldots, K$ with true signal $Z_k$ and error $E_k$, if $\{[V_k]\}_{k \in \mathcal{K} }$, the collection of $Z_k$'s signal score subspaces, is relatively independent, then, regardless of the ordering of index-sets,
% in $(\mathbbm{2}^{\mathcal{K} } \setminus  \{ \phi \}, \subset )$,
there exists a unique set of pairs $ \{ (S_i, r(S_i))  \}_{i \in \mathcal{I}_K } $.
\end{theorem}

Under only the relative independence condition, the determination of the partially-joint score subspaces $[W_i]$ corresponding to $S_i$, $i \in \mathcal{I}_K$, may not be unique, and depends on the ordering of index-sets  (see Examples A.3 and A.4 in the supplementary material). To ensure uniqueness of $[W_i]$'s, we require a rather strong assumption. We say that $\{ [V_k] \}_{k \in \mathcal{K}}$ is \textit{absolutely orthogonal}, if (1) $\{ [V_k] \}_{k \in \mathcal{K}}$ is relatively orthogonal, and (2) for each $l = 1, \ldots, K - 1$ and for every $i \in \mathcal{J}_l$,
\begin{align} \label{rule5}
P_{I_l}^\perp ( \cap_{k\in S_i} [V_{k}])= P_{J_i}^\perp ( \cap_{k\in S_i} [V_{k}]),
\end{align}
where $[J_i] = +_{j \in \mathcal{J}_{i,>l}}(\cap_{k\in S_j} [V_{k}]) = +_{j \in \mathcal{J}_{i,>l}} [W_j]$ and $\mathcal{J}_{i,>l} =  \{ j : |S_j| > l, S_i \cap S_k \neq \phi   \}$.

%{\color{red} In Example \ref{exam:examone}, when $l = 1$ and $i = 6$, (\ref{rule5}) holds if $ ( \bigcirc_{j \in \{1,2,3,4\}} P_j^\perp) ([V_1]) = ( \bigcirc_{j \in \{1,2,3\}} P_j^\perp) ([V_2])$. Note that for $i = 6$, $S_i = \{2\}$ and the index-sets $S_j$'s for $j \in \mathcal{J}_{i, >l}$ are $S_1 = \{1,2,3\}$, $S_2 = \{1,2\}$, $S_3 = \{2,3\}$, excluding $S_4 = \{ 1,3 \}$.}

\begin{theorem}\label{thm:thmtwo}
	Given matched data matrices $X_k = Z_k + E_k \in \mathbb{R}^{p_k \times n}$ for $k = 1, \ldots, K$ with true signal $Z_k$ and error $E_k$, if $\{[V_k]\}_{k \in \mathcal{K} }$ is absolutely orthogonal, then partially-joint score subspaces $[W_i]$ for $i \in \mathcal{I}_K$ are  uniquely determined.
\end{theorem}

The uniqueness of each partially-joint loading subspace $[U_{(k), i}]$ is deduced from the uniqueness of $[W_{i}]$'s.

\begin{corollary}\label{cor:u}
	Given matched data matrices $X_k = Z_k + E_k \in \mathbb{R}^{p_k \times n}$ for $k = 1, \ldots, K$ with true signal $Z_k$ and error $E_k$, if $\{[V_k]\}_{k \in \mathcal{K} }$ is absolutely orthogonal, then each partially-joint loading subspace $[U_{(k),i}]$ is uniquely determined for $k = 1, \ldots, K$ and $i \in \mathcal{I}_{(k)}$.
\end{corollary}

We provide proofs of theorems in this section and examples for relative independence and absolute orthogonality in Section A of the supplementary material.

\section{Simulation Study}
\label{sec:simul}

\subsection{Example Dataset Generation} \label{sec:exmample}

In the simulation study, we use the following data generation setting for numerically analyzing the performance of our proposal. Throughout, we use $K = 3$ blocks of data sets, in which the association structures are given by the ranks of index-sets.

First, we set a pre-determined rank $r(S_i)$ for each index-set $S_i$ for $i = 1, \ldots, 2^K - 1$.
 The partially-joint score matrix $W_{i} \in \mathbb{R}^{n \times r(S_i)}$ for each $S_i$ is given by the orthonormal basis of a random $n \times r(S_i)$ matrix consisting of independently standard normal variates.

The generic partially-joint loading matrix $U_{(k),i} \in \mathbb{R}^{p_k \times r(S_i)}$ for $i = 1, \ldots, 2^K - 1$ and $k = 1, \ldots, K$ is a column-wise concatenation of randomly generated vectors $u_{k,i,j} \in \mathbb{R}^{p_k}$, $j = 1, \ldots, r(S_i)$. Each entry of $u_{k,i,j}$ is generated following $\mathcal{N}(0, \sigma_{i,j}^2)$ independently. The magnitude of signal $\sigma_{i,j}^2$'s depend on the simulation settings, and are summarized as $\sigma_{M}^2 =  \{ ( \sigma_{i,j}^2 )  : i  = 1,\ldots, 2^{K-1}, j = 1, \ldots, r(S_i),  r(S_i) > 0 \}$.

We then derive the generic signal matrix $Z_k$ for $k = 1 \ldots, K$ by $Z_k = \sum_{i=1}^{2^K - 1} U_{(k), i} W_{ i}^T $. The concatenation of generic loading matrices is denoted $U$, and is of size $\sum_{k=1}^K p_k \times \sum_{i=1}^{2^K-1} r(S_i)$.

The error matrix $E_k$ is generated element-wisely, such that $e_{k, ij} \sim \mathcal{N}(0, \sigma^2)$ independently for $i = 1, \ldots, p_k$ and $j = 1, \ldots, n$. The magnitude of error $\sigma^2$ is set as the reciprocal of the signal-to-noise ratio (SNR),
%$\sigma^2 = 1 / ( c \cdot \textrm{SNR})$, {\color{red} where $c = n / 200$ is a scaling factor. SNR is predetermined as a simulation setting. [Is the number 200 appear for other choices of $n \neq 200$?]}
$\sigma^2 = 1 / ( \textrm{SNR})$. % {\color{red} where $c = n / 200$ is a scaling factor. SNR is predetermined as a simulation setting. [Is the number 200 appear for other choices of $n \neq 200$?]}

We use the following six models.
\begin{enumerate}
	\item (Individuals) $\mathfrak{S} = \{ (\{ 1 \}, 2 ), (\{ 2 \}, 2), (\{ 3 \}, 2) \}$,  $\sigma_{M}^2 = \{ (1.4, 0.8), (1.3, 0.7), \linebreak (1.2, 0.6)  \}$
	\item (Fully joint) $\mathfrak{S} = \{ (\{ 1 ,2 ,3 \},2) \}$,  $\sigma_{M}^2 = \{ (1.0, 0.9) \}$
	\item (Circular, partially joint) $\mathfrak{S} = \{ (\{ 1 ,2 \},2),(\{ 1 , 3\},2) , (\{ 2 , 3\},2) \}$,  $\sigma_{M}^2 = \{ (1.4, 0.8),  (1.3,0.7), (1.2,0.6) \}$
	\item (Mix of fully joint and individuals)  $\mathfrak{S} = \{ (\{ 1 ,2, 3 \},2), (\{ 1 \},2) , (\{ 2 \},2),\linebreak (\{ 3 \},2) \}$,  $\sigma_{M}^2 = \{ (1.5, 0.8),  (1.4,0.7), (1.3,0.6),(1.2,0.5) \}$
	\item (Fully joint and partially joint)  $\mathfrak{S} = ( \{ 1 , 2, 3\},2 ), ( \{ 1, 2 \},2),  (\{ 1 ,3 \},2), \linebreak ( \{ 2,3 \} ,2) \}$,  $\sigma_{M}^2 = \{ (1.5, 0.8), (1.4,0.7), (1.3,0.6),(1.2,0.5) \}$
	\item (All possible combinations)   $\mathfrak{S} = ( \{ 1 , 2, 3\} ,2 ) ,  ( \{ 1, 2 \} , 2 ) , ( \{ 1, 3 \} , 2), \linebreak (\{ 2,3 \}, 2),  ( \{ 1 \} ,2), ( \{ 2 \},2), (\{ 3 \},2) \}$,  $\sigma_{M}^2 = \{ (1.8, 0.8), (1.7,0.7), (1.6,0.6),\linebreak (1.5,0.5),(1.4,0.4),  (1.3,0.3),(1.2,0.2) \}$
\end{enumerate}

\subsection{Results on Comparative Study} \label{sec:sec5.3}

In this subsection, we numerically compare the performance of our proposal to other competitors, including SLIDE \citep{Gaynanova2019}, COBS \citep{Gao2020}, AJIVE \citep{Feng2018}, and JIVE \citep{Lock2013}. From the estimates of each method, the partially-joint structure $\mathfrak{S}$, as well as the concatenated partially-joint score matrix $\widehat{W}$ and loading matrix $\widehat{U}$  can be extracted. See Section B.2 of the supplementary material for a brief review of these methods. Our proposal will be called the method of partially-joint structure identification, or PSI for short.

% measures of quality

To assess the efficacy of finding the true partially-joint structure ${\mathfrak{S}}$ and proper loading and score matrices, we use the following measures for PSI and other four methods.

\begin{enumerate}
	\item[(1)] \textit{Accuracy}:
	The rate of finding the true partially-joint structure, $\mathbb{E} \mathbbm{1} (\mathfrak{S} = \widehat{\mathfrak{S}})$.

        \item[(2)]  \textit{Mean relative squared error (RSE)}: The relative size of the reconstruction error for each $Z_k$,
        \begin{align*}
            \frac{1}{K} \sum_{k=1}^K \frac{\Vert Z_k - \widehat{U}_{(k)} \widehat{W}_{(k)}^T \Vert_F^2}{\Vert Z_k \Vert_F^2}.
        \end{align*}
	\item[(3)] $\overline{\theta}(U,\widehat{U})$ for \textit{Partially-joint loading matrix} $\widehat{U}$:
	We measure the difference between $U$ and $\widehat{U}$ as follows. We denote the principal angles between $U_{k,i}$ and $\widehat{U}_{(k)}$ by $\theta_{U,k,i,j}$ for $k = 1, \ldots, K$, $i = 1, \ldots, 2^K-1$ and $j = 1, \ldots, r(S_i)$. We report the average $\overline{\theta}(U,\widehat{U})$ of all $\theta_{U,k,i,j}$'s.
	
 \item[(4)] $\overline{\theta}(W,\widehat{W})$ for \textit{Partially-joint score} $\widehat{W}$: We measure the difference between $W$ and $\widehat{W}$ as follows. We denote the principal angle between $w_{i,j}$ and $\widehat{W}$ by $\theta_{W,i,j}$ for $i = 1, \ldots, 2^K-1$ and $j = 1, \ldots, r(S_i)$. We report the average $\overline{\theta}(W,\widehat{W})$ of all $\theta_{W,i,j}$'s as .
	
\end{enumerate}

The simulation was conducted on different values of SNR ($10$ and $5$) for the example models $1$ to $6$ with $n = 200$ and $p_1 = p_2 = p_3 = 200$. For each model, we randomly generate a true partially-joint loading matrix $U$ which is then fixed throughout repetitons. For each SNR value, we repeat the simulation 100 times. %then the data generating procedures Given a fixed true partially-joint loading matrix $U$, $100$ data sets were generated for each SNR value.

We performed a comparative study of all 5 methods over these datasets. The initial ranks were given to each method as follows.
For PSI and AJIVE, we have given the initial ranks for each $X_k$ by their own true rank. For example, in Model 3, $r_k = 4$ for each $k$.
SLIDE and JIVE require the concatenated data matrix $X$ as input, thus we have used the whole data matrix for these methods.   (We also have used reduced-rank matrices, with the ranks given by the true model, as input for SLIDE and JIVE, but the results turned out to be worse.)
For COBS, we use the concatenated $X$ and set the number of components as the sum of each index-set's true rank (for example, $6 = 2+2+2$ in Model 3).
%
%(1) PSI, AJIVE : truncate each $X_k$ with its own true rank (for example, $4$ in Model 3),  (2) SLIDE, JIVE : input row-concatenation $X$ of $X_k$, (3) COBS : input the whole row-wise concatenated $X_k$'s and set the number of components as the sum of each index-set's true rank (for example, $6 = 2+2+2$ in Model 3).
%
Accuracy of finding the true structure and the average (and standard deviation) of  RSE, $\overline{\theta}(U,\widehat{U})$ and $\overline{\theta}(W,\widehat{W})$, over 100 repetitions, are reported in Tables \ref{tab:tabone} and \ref{tab:tabtwo}.

\begin{table}[hbt!]
	\caption{Comparative Study on Model 1 to 3. The unit for Accuracy is percent.  \label{tab:tabone}}
	\scriptsize
	\centering
	\resizebox{\textwidth}{!}{%
		\begin{tabular}{cccccccc}
			\hline
			Model & SNR & Measure & PSI & SLIDE bcv & COBS & AJIVE & JIVE \\
			\hline \hline
1 & 15 & Accuracy   & \textbf{100} & \textbf{100}  & 30  & 96  & 91  \\
  &  & RSE & \textbf{0.24 (0.01)} & \textbf{0.24 (0.01)} & 0.41 (0.04) & \textbf{0.24 (0.02)} & 0.26 (0.03) \\
  &  &  $\overline{\theta}(U,\widehat{U})$   & \textbf{15.99 (0.47)} & 50.99 (1.59) & 54.84 (5.77) & 16.04 (0.59) & 16 (0.48) \\
  &  &  $\overline{\theta}(W,\widehat{W})$      & \textbf{21.18 (0.39)} & \textbf{21.18 (0.4)} & 23.97 (0.84) & 21.45 (1.27) & 21.22 (0.4) \\  \cline{2-8}
 & 10 & Accuracy   & \textbf{100}  & 85  & 47  & 97  & 82  \\
  &  & RSE & \textbf{0.38 (0.01)} & 0.39 (0.03) & 0.56 (0.04) & \textbf{0.38 (0.03)} & 0.42 (0.09) \\
  &  &  $\overline{\theta}(U,\widehat{U})$   & \textbf{19.93 (0.66)} & 49.32 (10.35) & 60.77 (4.87) & 19.99 (0.74) & \textbf{19.93 (0.68)} \\
  &  &  $\overline{\theta}(W,\widehat{W})$      & \textbf{25.94 (0.59)} & 27.22 (3.38) & 31.4 (1.99) & 26.15 (1.1) & 25.99 (0.59) \\  \cline{2-8}
 & 5 & Accuracy   & \textbf{100}  & 0  & 26  & 95  & 34  \\
  &  & RSE & 0.89 (0.03) & 0.87 (0.03) & \textbf{0.85 (0.04)} & 0.9 (0.03) & 1.17 (0.33) \\
  &  &  $\overline{\theta}(U,\widehat{U})$   & 30.22 (1.06) & \textbf{25.45 (9.27)} & 66.16 (5.21) & 30.38 (1.23) & 29.98 (1.02) \\
  &  &  $\overline{\theta}(W,\widehat{W})$      & 36.58 (1.02) & 65.48 (3.75) & 49.3 (2.46) & 37.1 (2.22) & \textbf{36.4 (0.98)} \\
  \hline
2 & 15 & Accuracy   & \textbf{100}  & \textbf{100} & \textbf{100}  & \textbf{100}  & 61  \\
   & & RSE & \textbf{0.13 (0.01)} & \textbf{0.13 (0.01)} & 0.22 (0.02) & 0.14 (0.01) & 0.19 (0.09) \\
   & &  $\overline{\theta}(U,\widehat{U})$   & 15.6 (0.46) & 15.58 (0.46) & 24.63 (1.84) & 15.6 (0.46) & \textbf{15.53 (0.47)} \\
   & &  $\overline{\theta}(W,\widehat{W})$      & 13.17 (0.48) & 12.83 (0.48) & 13.52 (0.54) & 13.19 (0.48) & \textbf{12.81 (0.47) }\\  \cline{2-8}
 & 10 & Accuracy   & \textbf{100} & \textbf{100} & \textbf{100} & \textbf{100} & 48 \\
  &  & RSE & 0.21 (0.01) & \textbf{0.2 (0.01)} & 0.36 (0.03) & 0.21 (0.01) & 0.31 (0.15) \\
  &  &  $\overline{\theta}(U,\widehat{U})$   & 19.13 (0.62) & 19.06 (0.63) & 32.28 (2.11) & 19.13 (0.62) & \textbf{18.99 (0.62)} \\
  &  &  $\overline{\theta}(W,\widehat{W})$      & 16.36 (0.63) & 15.85 (0.62) & 17.35 (0.8) & 16.4 (0.63) & \textbf{15.8 (0.62)} \\  \cline{2-8}
 & 5 & Accuracy   & 61 & 3 & 36 & \textbf{99} & 62 \\
  &  & RSE & 0.62 (0.19) & 0.63 (0.11) & 0.74 (0.09) & \textbf{0.49 (0.03)} & 0.6 (0.19) \\
  &  &  $\overline{\theta}(U,\widehat{U})$   & 28.64 (1.97) & 28.55 (7.79) & 50.61 (7.39) & 27.51 (0.95) & \textbf{27.08 (0.95)} \\
  &  &  $\overline{\theta}(W,\widehat{W})$      & 24.25 (0.99) & 25.89 (2.38) & 32.38 (7.46) & 24.55 (0.98) & \textbf{23.13 (0.88)} \\
\hline
3 & 15 & Accuracy   & \textbf{100} & \textbf{100} & 0  & 0  & 0  \\
  &  & RSE & \textbf{0.16 (0)} & \textbf{0.16 (0.01)} & 0.35 (0.02) & 0.27 (0.01) & 0.48 (0.13) \\
  &  &  $\overline{\theta}(U,\widehat{U})$   & 15.34 (0.38) & 29.99 (3.98) & 29.16 (3.14) & 16.01 (0.45) & \textbf{14.88 (0.39)} \\
  &  &  $\overline{\theta}(W,\widehat{W})$      & 15.56 (0.35) & 15.23 (0.33) & 17.08 (0.5) & 16 (0.57) & \textbf{15.06 (0.4)} \\  \cline{2-8}
 & 10 & Accuracy   & \textbf{100}  & \textbf{100}  & 2  & 0  & 0  \\
   & & RSE & 0.25 (0.01) & \textbf{0.24 (0.01)} & 0.52 (0.04) & 0.4 (0.01) & 0.74 (0.16) \\
  &  &  $\overline{\theta}(U,\widehat{U})$   & 18.91 (0.44) & 32.74 (4.79) & 41.14 (4.44) & 20.3 (0.68) & \textbf{18.34 (0.47)} \\
  &  &  $\overline{\theta}(W,\widehat{W})$      & 19.28 (0.43) & 18.79 (0.39) & 23.25 (1.31) & 20.67 (1.05) & \textbf{18.69 (0.43)} \\  \cline{2-8}
 & 5 & Accuracy   & \textbf{10}  & 0  & 0 & 0  & 0 \\
   & & RSE & 0.82 (0.11) & 0.79 (0.07) & 0.84 (0.03) & \textbf{0.78 (0.03)} & 1.8 (0.29) \\
   & &  $\overline{\theta}(U,\widehat{U})$   & 29.64 (1.22) & 54.59 (9.9) & 67.45 (5.18) & 29.9 (1.02) & \textbf{26.79 (0.75)} \\
   & &  $\overline{\theta}(W,\widehat{W})$      & 27.52 (0.72) & 31.96 (3.58) & 43.38 (2.76) & 31.47 (2.03) & \textbf{27.01 (0.78)} \\  \hline
			
		\end{tabular}
	}
\end{table}

In Model $1$, in which case there are only individual scores, PSI and AJIVE have identified the true partially-joint structure for almost all instances, but the performance of PSI is slightly better.
% Note that AJIVE finds the true structure under rate of 95\%.
SLIDE is comparable in terms of Accuracy and RSE and but sometimes provides large deviations in estimating the partially-joint loadings $U$.
%
% partially-joint scores well as PSI and AJIVE, but shows large values in finding partially-joint loadings.
One possible reason is that SLIDE estimates partially-joint scores by enforcing $W^T W = I_{\widehat{r} \times \widehat{r}}$, whereas PSI and AJIVE allow non-orthogonality of these scores between different index-sets.

In Model $2$, in which case there are only joint score, all methods but JIVE find the true structure quite well for large enough SNRs, and PSI, SLIDE and AJIVE have estimated almost identical partially-joint score subspaces. %SLIDE finds identical partially-joint loadings as PSI and AJIVE in this case, as model 2 is equivalent to the truncated SVD of the row-concatenation matrix of $X_k$'s.

In Model $3$, in which partially-joint scores are entangled in a cyclic structure, PSI and SLIDE show good performance in identifying the true structures. On the other hand, other methods have completely failed to identify the partially joint structure, as expected.
We observe that AJIVE and JIVE yield surprisingly nice figures in
$\overline{\theta}(U,\widehat{U})$ and
$\overline{\theta}(W,\widehat{W})$, and it may call an explanation. For Model 3, AJIVE and JIVE estimate a fully-joint structure rather than the (true) partially-joint structure. In spite of this inaccurate identification of structures, the estimation of $U$ and $W$ for AJIVE and JIVE becomes similar to applying an SVD for the concatenated $X$, which sometimes yields a good performance.
%
%
% as a multiple set of fully-joint structure, \textit{e.g.} $\{ (\{1,2,3\},6) \}$, as like the SVD of $X$, the whole row-concatenation of data blocks $X_k$'s, trying to minimize the overall reconstruction error.
% In  cases like this, the column spaces spanned by $\widehat{U}$ and $\widehat{W}$ estimate the left and right singular vector spaces of $X$ with a large rank.
%  Then $\widehat{U}$ and $\widehat{W}$ are actually linear subspace quite large so as close as to the whole left and right singular vector spaces of $X$ in $\mathbb{R}^n$ respectively. As the method finds $\widehat{U}$ and $\widehat{W}$ large as possible, $\overline{\theta}(U,\widehat{U})$  $\overline{\theta}(W,\widehat{W})$ are estimated low as much. Table \ref{tab:tabone} confirms these observations.

Simulation results for Models 4--6 are contained in Table~\ref{tab:tabtwo}.
Model $4$ represents the case where both joint and individual scores are present.
Both SLIDE and AJIVE showed superior performance in identifying the true structure in this case, and PSI was as competent in estimating score subspaces as AJIVE. Models $5$ and $6$ are most complicated with all types of associations (joint, partially-joint and individual) are present. For these cases both PSI and SLIDE show superior performances than other methods.
(For Model 6, the algorithm for COBS did not converge.) Overall, the proposed PSI performs best in terms of the accuracy in estimating the true structure, for all models considered, and is comparable to the competitors in terms of RSE and estimating the loading and score subspaces. Among the competitors, we find SLIDE's overall performance is among the best. While JIVE surprisingly shows great performance in estimating the loading and score subspaces, we emphasize that JIVE fails in identifying the block-wise association structures, measured by the accuracy, for almost all models.

 %scores mixed, PSI and SLIDE is prominent in estimating true structure. See Table~\ref{tab:tabtwo}.

\begin{table}[hbt!]
	\caption{Comparative Study on Model 4 to 6. The unit for Accuracy is percent.    \label{tab:tabtwo}}
	\scriptsize
	\centering
	\resizebox{\textwidth}{!}{%
		\begin{tabular}{cccccccc}
			\hline
			Model & SNR & Measure & PSI & SLIDE bcv & COBS & AJIVE & JIVE \\
			\hline \hline
4 & 15 & Accuracy   & \textbf{100}  & \textbf{100}  & 15  & \textbf{100}  & 98  \\
  &  & RSE & 0.18 (0.01) & \textbf{0.17 (0.01)} & 0.33 (0.03) & \textbf{0.17 (0)} & 0.19 (0.01) \\
  &  &  $\overline{\theta}(U,\widehat{U})$   & 15.11 (0.31) & 28.76 (6.35) & 35.5 (4.08) & \textbf{15.09 (0.31)} & 15.27 (0.31) \\
  &  &  $\overline{\theta}(W,\widehat{W})$      & 18.97 (0.38) & \textbf{18.72 (0.37)} & 21.65 (0.86) & 18.84 (0.37) & 18.78 (0.38) \\  \cline{2-8}
 & 10 & Accuracy   & \textbf{100}  & 95  & 53  & \textbf{100}  & 69  \\
  &  & RSE & \textbf{0.28 (0.01)} & \textbf{0.28 (0.01)} & 0.47 (0.03) & \textbf{0.28 (0.01)} & 0.31 (0.03) \\
  &  &  $\overline{\theta}(U,\widehat{U})$   & 18.82 (0.45) & 32.05 (5.6) & 43.66 (3.55) & \textbf{18.79 (0.45)} & 18.87 (0.47) \\
  &  &  $\overline{\theta}(W,\widehat{W})$      & 23.28 (0.51) & 23.13 (1.18) & 28.58 (1.73) & 23.1 (0.49) & \textbf{22.99 (0.49)} \\  \cline{2-8}
 & 5 & Accuracy   & 24  & 5  & 4  & \textbf{100}  & 15  \\
  &  & RSE & 0.75 (0.08) & 0.71 (0.07) & 0.76 (0.03) & \textbf{0.64 (0.02)} & 0.89 (0.18) \\
  &  &  $\overline{\theta}(U,\widehat{U})$   & 28.46 (0.95) & 55.1 (10.3) & 58.55 (4.85) & 27.56 (0.66) & \textbf{27.22 (0.7)} \\
  &  &  $\overline{\theta}(W,\widehat{W})$      & 32.87 (0.87) & 42.97 (7.98) & 48.73 (3.46) & 33.04 (0.8) & \textbf{32.36 (0.83)} \\
   \hline
5 & 15 & Accuracy   & \textbf{100}  & \textbf{100}  & 0  & 0  & 0  \\
  &  & RSE & \textbf{0.14 (0)} & \textbf{0.14 (0)} & 0.3 (0.02) & 0.2 (0.01) & 0.47 (0.08) \\
  &  &  $\overline{\theta}(U,\widehat{U})$   & 14.78 (0.29) & 21.89 (4.68) & 26.81 (1.32) & 15.09 (0.32) & \textbf{14.12 (0.4)} \\
  &  &  $\overline{\theta}(W,\widehat{W})$      & 14.93 (0.3) & 14.31 (0.29) & 16.06 (0.38) & 14.92 (0.38) & \textbf{13.77 (0.38)} \\  \cline{2-8}
 & 10 & Accuracy   & \textbf{100}  & 99  & 0  & 0  & 0  \\
  &  & RSE & 0.23 (0.01) & \textbf{0.22 (0.01)} & 0.46 (0.02) & 0.31 (0.01) & 0.73 (0.1) \\
  &  &  $\overline{\theta}(U,\widehat{U})$   & 18.3 (0.41) & 26.21 (3.95) & 36.22 (2.49) & 18.88 (0.44) & \textbf{17.47 (0.51)} \\
  &  &  $\overline{\theta}(W,\widehat{W})$      & 18.52 (0.35) & 17.67 (0.33) & 22.04 (1.3) & 18.83 (0.62) & \textbf{17.03 (0.43)} \\  \cline{2-8}
 & 5 & Accuracy   & \textbf{9}  & 0  & 0  & 0  & 0  \\
   & & RSE & 0.75 (0.11) & 0.68 (0.05) & 0.75 (0.03) & \textbf{0.66 (0.03)} & 1.58 (0.18) \\
   & &  $\overline{\theta}(U,\widehat{U})$   & 28.64 (1.14) & 50.51 (6.92) & 57.92 (3.66) & 27.97 (0.73) & \textbf{25.28 (0.78)} \\
   & &  $\overline{\theta}(W,\widehat{W})$      & 26.18 (1.1) & 29.69 (2.18) & 39.35 (2.21) & 29.23 (1.44) & \textbf{24.73 (0.7)} \\
\hline
6 &15 & Accuracy   & \textbf{100}  & 96 & -  & 0 & 0  \\
  &  & RSE & \textbf{0.15 (0)} & \textbf{0.15 (0)} & - & 0.17 (0.01) & 0.19 (0.02) \\
  &  &  $\overline{\theta}(U,\widehat{U})$   & 14.44 (0.26) & 34.21 (2.7) & - & \textbf{14.3 (0.27)} & 14.35 (0.42) \\
  &  &  $\overline{\theta}(W,\widehat{W})$      & 17.5 (0.3) & \textbf{16.99 (0.27)} & - & 17.3 (0.3) & 17.09 (0.29) \\  \cline{2-8}
 & 10 & Accuracy   & \textbf{100}  & 73 & - & 0 (0) & 0 (0) \\
  &  & RSE & \textbf{0.24 (0)} & \textbf{0.24 (0.01)} & - & 0.27 (0.01) & 0.36 (0.05) \\
  &  &  $\overline{\theta}(U,\widehat{U})$   & 17.92 (0.29) & 35.78 (3.15) &- & 17.82 (0.36) & \textbf{17.63 (0.53)} \\
  &  &  $\overline{\theta}(W,\widehat{W})$      & 21.53 (0.31) & 21.53 (1.62) & - & 21.33 (0.38) & \textbf{20.77 (0.4)} \\  \cline{2-8}
 & 5 & Accuracy   & \textbf{18}  & 0  & - & 0  & 0 \\
  &  & RSE & 0.66 (0.07) & \textbf{0.63 (0.03)} & - & 0.65 (0.02) & 1.14 (0.11) \\
  &  &  $\overline{\theta}(U,\widehat{U})$   & 27.35 (0.77) & 52.93 (6.03) & -  & 27.18 (0.55) & \textbf{25.24 (0.69)} \\
  &  &  $\overline{\theta}(W,\widehat{W})$      & 30.16 (0.97) & 37.01 (4.35) & - & 31.72 (0.82) & \textbf{28.54 (0.65)} \\ \hline
		\end{tabular}
	}
\end{table}

We also have compared the computation time needed by each method. Our method, PSI, took remarkably small computation times in all situations. For each of six models, PSI took 0.8---9.1 seconds in computation, compared to the time ranges 240---940 seconds for SLIDE, 10---73 seconds for COBS (when converged) and 6.3---417 seconds for JIVE. AJIVE took about 12---30 seconds. Table B.5  in the supplementary material lists the computation time for our method and other four methods.

\subsection{Results on Imbalanced Signal Strength between Joint and Individual Components}

Next, we consider cases where the signal strengths of joint components and individual components are grossly imbalanced. Consider a new model, with $K = 3$ and $p_1 = p_2 = p_3 = 100$, whose index-set ordering is $S_1 = \{1,2,3\}$, $S_5 = \{1\}$, $S_6 = \{2\}$ and $S_7 = \{3\}$. We set inherent joint rank $r(S_1) = 10$ and individual ranks $r(S_5) = r(S_6) = r(S_7) = 10$. Other index-sets have zero ranks, $r(S_2) = r(S_3) = r(S_4) = 0$. Throughout, $n = 200$.

For the case in which there are larger variations in the joint component than the individual components, we set $\sigma_{1,j}^2 \gg \sigma_{i,j}^2$ for $i = 5,6,7$ and all $j$'s. In the opposite case, we give larger variations in the individual components than the joint component, $\sigma_{i,j}^2 \gg \sigma_{1,j}^2$ for $i = 5,6,7$ and all $j$'s. Details are given in Section B.5 of the supplementary materials.

We carried out comparative simulations on both cases at SNR levels $\infty$ and $5$. Not only the success rate of finding true structure but also the numbers of estimated joint and individual components were evaluated. In Table \ref{tab:tabthree}, in which case the joint component has larger variations, PSI and JIVE showed good performances to other methods. SLIDE failed to identify (weak) individual structures. AJIVE showed a compatible performance at SNR 10. When the individual components have larger variations, in Table \ref{tab:tabfour}, AJIVE is the best to detect (weak) joint structure, boasting the capability of Wedin bound simulation. SLIDE and JIVE also show compatible performances.

%% Unbalanced Data (Joint)

\begin{table}[bt]
	\caption{The case where the joint component has larger variations than the individual components.\label{tab:tabthree}}
	\centering
	\scriptsize
	\resizebox{\textwidth}{!}{%
		\begin{tabular}{ccccccc}
			\hline
			SNR & Measure & PSI & SLIDE bcv & COBS & AJIVE & JIVE \\
			\hline
			$\infty$ & Accuracy   & \textbf{100}  & 0  & 0  & 0  & \textbf{100}  \\
			& Joint & \textbf{10 (0)} & \textbf{10 (0)} & 39.62 (0.62) & 4.03 (2.12) & \textbf{10 (0)} \\
			& Individual &  \textbf{30 (0)} & 0 (0) & 0.04 (0.2) & 47.91 (6.36) & \textbf{30 (0)} \\  \hline
			10 & Accuracy & \textbf{100}  & 0  & 0  & 0  & 9  \\
			& Joint &  \textbf{10 (0)} & \textbf{10 (0)} & 14.01 (1.63) & \textbf{10 (0)} & \textbf{10 (0)} \\
			& Individual & \textbf{30 (0)} & 0 (0) & 17.14 (2.67) & 32.99 (0.85) & 34.07 (4.65) \\
			\hline
	\end{tabular} }
\end{table}

%% Unbalanced Data (Individual)
\begin{table}[bt]
	\caption{The case where the individual components have larger variations than the joint components. \label{tab:tabfour}}
	\centering
	\scriptsize
	\resizebox{\textwidth}{!}{%
		\begin{tabular}{ccccccc}
			\hline
			SNR & Measure & PSI & SLIDE bcv & COBS & AJIVE & JIVE \\
			\hline
			$\infty$ & Accuracy & \textbf{100}  & 0  & 0  & 0  & 0  \\
			& Joint & \textbf{10 (0)} & 10.47 (0.73) & 38.95 (1.06) & 3.8 (1.83) & 14.42 (4.05) \\
			& Individual & \textbf{30 (0)} & 25.81 (1.41) & 0.09 (0.29) & 48.6 (5.48) & 27.91 (1.17) \\ \hline
			10 & Accuracy & 0  & 0  & 0  & \textbf{100 } & 0  \\
			& Joint & 1.56 (2.15) & 10.79 (1.27) & 29.61 (2.89) & \textbf{10 (0)} & 13.34 (3.66) \\
			& Individual & 46.84 (9.15) & 23.37 (1.76) & 3.45 (1.5) & \textbf{30 (0)} & 27.09 (1.19) \\
			\hline
	\end{tabular} }
\end{table}

\section{Real Data Analysis}
\label{sec:realdata}

In this section, we apply the proposed  PSI  to a blood cancer multi-omics data set linked to a drug response panel \citep{Dietrich2018}, available in the European Genome-Phenome Archive  under accession number {\texttt{EGAS0000100174}}. We have chosen to use 121 cases diagnosed with chronic lymphocytic leukemia (CLL). The drug response panel ($X_{\textrm{Drug}}$) records the \textit{ex vivo} cell viabilities at a series of 5 concentrations, for each of 62 drugs that target onco-related pathways or are used widely in clinical practice.
This multi-omics data set consists of the genome-wide DNA methylation profiles and the RNA sequencing profiles. The top 5,000 most variable CpG sites were selected ($X_{\textrm{Meth}}$) from the \textit{450K illumina assay} DNA methylation profile. As for the RNA sequencing profile, we selected the top 5,000 gene expressions with the largest stabilized variances ($X_{\textrm{Exp}}$) from the high-throughput sequencing (HTS) assay. In summary, we have three blocks of  data sets  $X_{\textrm{Drug}} \in \mathbb{R}^{310 \times 121}$, $X_{\textrm{Meth}} \in \mathbb{R}^{ 5000 \times 121}$ and $X_{\textrm{Exp}} \in \mathbb{R}^{5000 \times 121}$.

We repeat the estimation process over $100$ repetitions (of data splitting) and select the mode structure, that is, the estimated partially-joint structure that appears the most out of $100$ repetitions. The estimated mode structure is
\begin{align*}
 \widehat{\mathfrak{S}} =  \{ (\{\textrm{Drug},{\textrm{Meth}}\}, 1) , (\{\textrm{Drug}\},4) , (\{{\textrm{Meth}}\},41) , (\{ {\textrm{Exp}} \},3)   \},
\end{align*}
which means that PSI detected the index-set $\{\textrm{Drug},{\textrm{Meth}}\}$ of rank $1$, and no fully-joint score or other forms of partially-joint scores were detected. This structure appeared 53 times out of 100 repetitions. The index-set $\{\textrm{Drug},{\textrm{Meth}}\}$ of rank 1 stands for the existence of one-dimensional latent score, partially-joint for Drug and Meth data sets (but not for Exp data set).

PSI showed better performance in computation time over other methods. In an Intel \textsuperscript{\textregistered} Xeon \textsuperscript{\textregistered} CPU E5-2640 v4 @ 2.40GHz system, it took 9.38 seconds per single data splitting instance (54.56 seconds for 100 instances with 40 cores). In comparison, the SLIDE with bcr method detected the same structure as our method, but it took almost 9 hours on the same machine. The COBS yielded an eccentric result, giving the fully-joint score of rank 50, taking about 14 minutes.

PSI was also robust over different choices of initial ranks. As a preprocessing, we set the ranks of the signal matrices using the principal component analysis which accounts for cumulative proportion of variances at $30\%$, $40 \%$, $50 \%$ and $60 \%$. Resulting signal ranks for   Drug, Meth and Exp datasets are $(2,6,1)$, $(2,15,1)$, $(3,28,2)$ and $(5,42,3)$, respectively. For all cases, PSI detected the same index-set $\{\textrm{Drug},{\textrm{Meth}}\}$ of rank $1$.

\begin{figure}[bt]
	\begin{center}
		\includegraphics[width=6in]{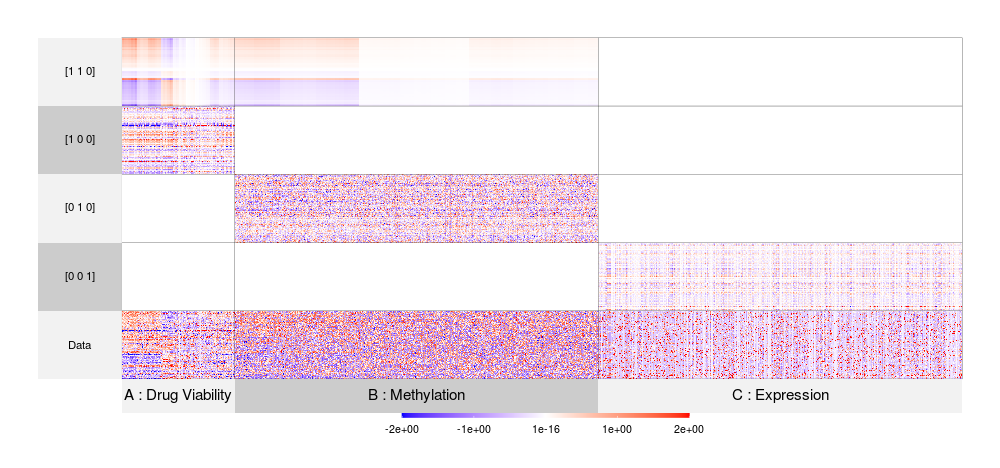}
	\end{center}
	\caption{The reconstructed components according to the estimated partially-joint structure. The values are all normalized for each column, then truncated to line on $[-2,2]$. \label{fig:third}}
\end{figure}

The analysis by PSI for the patient sample multi-block data reveals a particular pattern in the latent partially-joint score, which cannot be identified by applying, e.g. principal component analysis to each data block.
%captured a latent partially-joint   that separate PCA analysis on a single data block cannot identify.
To support this claim, we plot the reconstructed matrices $\widehat{U}_{(k),i}\widehat{W}_i^T$ for the identified partially-joint and individual parts of the data in Fig.~\ref{fig:third}.
Both the samples and the variables of $X_{\textrm{Drug}}$ are ordered by a hierarchical bi-clustering applied to the partially-joint component $\widehat{U}_{(\textrm{Drug}),2}\widehat{W}_2^T$ of $X_{\textrm{Drug}}$, where $\widehat{W}_2 = \widehat{W}_{\{\textrm{Drug},{\textrm{Meth}}\}}$.
The matrix $Z_{\textrm{Drug},S_2}:= \widehat{U}_{(\textrm{Drug}),2}\widehat{W}_2^T$ is shown in the top left part of Fig.~\ref{fig:third}.
  The variables of $X_{\textrm{Meth}}$ are similarly ordered.

%  $\{ 1, 1, 0 \}$, denoted $X_{\textrm{Drug}}^{110}$, shown in the top left part of Figure \ref{fig:third}.

\begin{figure}[bt]
	\centering
	\begin{subfigure}[b]{0.45\textwidth}
		\centering
		\subcaption{$Z_{\textrm{Drug},S_2}$ \label{fig:4a}}
		\includegraphics[width=\textwidth]{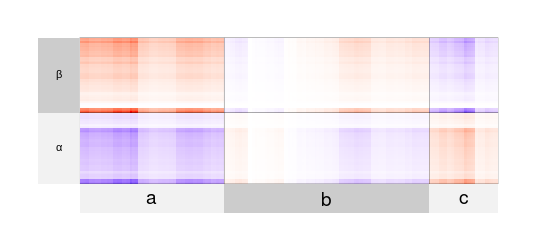}
	\end{subfigure}
	\begin{subfigure}[b]{0.45\textwidth}
		\centering
		\subcaption{the whole $X_{\textrm{Drug}}$ \label{fig:4b}}
		\includegraphics[width=\textwidth]{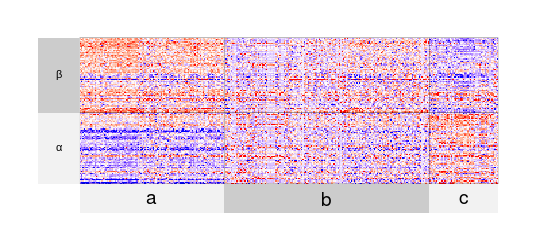}
	\end{subfigure}

	\begin{subfigure}[b]{0.45\textwidth}
		\centering
		\subcaption{$Z_{\textrm{Drug},S_2}$ \label{fig:4c}}
		\includegraphics[width=\textwidth]{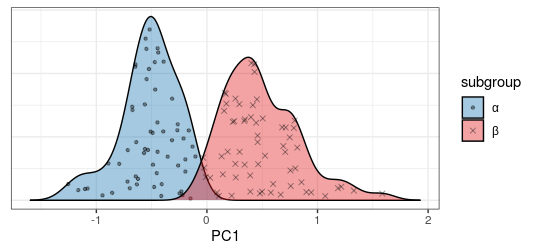}
	\end{subfigure}
	\begin{subfigure}[b]{0.45\textwidth}
		\centering
		\subcaption{the whole $X_{\textrm{Drug}}$ \label{fig:4d}}
		\includegraphics[width=\textwidth]{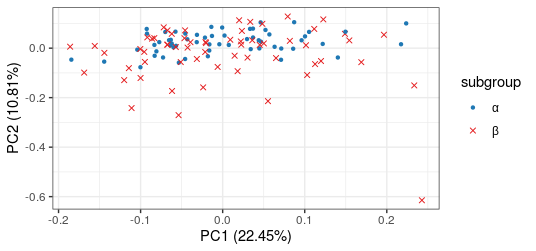}
	\end{subfigure}

	\begin{subfigure}[b]{0.45\textwidth}
		\centering
		\subcaption{$Z_{\textrm{Meth},S_2}$ \label{fig:4e}}
		\includegraphics[width=\textwidth]{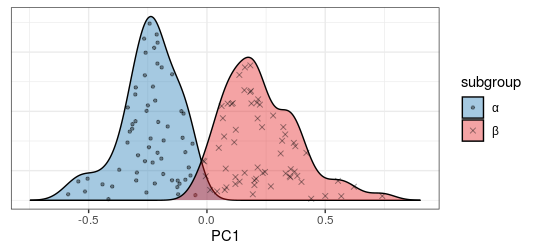}
	\end{subfigure}
	\begin{subfigure}[b]{0.45\textwidth}
		\centering
		\subcaption{the whole $X_{\textrm{Meth}}$ \label{fig:4f}}
		\includegraphics[width=\textwidth]{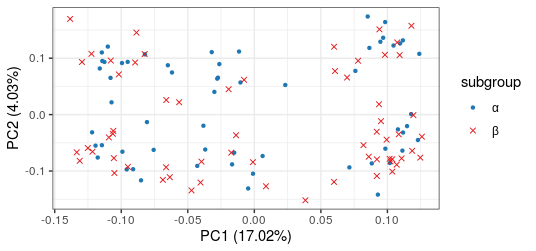}
	\end{subfigure}
	
	\caption{(a) The matrix $Z_{\textrm{Drug},S_2}:= \widehat{U}_{(\textrm{Drug}),2}\widehat{W}_2^T$.
(b) The whole $X_{\textrm{Drug}}$ data matrix.
(c) The density plot of the subgroups $\alpha$ and $\beta$ along the first principal component (PC) score of $Z_{\textrm{Drug},S_2}$; (d) The PC scores plot for the whole $X_{\textrm{Drug}}$; (e) The density plot of the subgroups $\alpha$ and $\beta$ along the first principal component of $Z_{\textrm{Meth},S_2}$; (f) The PC scores plot of the whole $X_{\textrm{Meth}}$.\label{fig:fourth}}
\end{figure}

Focusing on the partially-joint scores corresponding to $\{\textrm{Drug},{\textrm{Meth}}\}$, the samples are clustered into two distinct subgroups. These subgroups are shown in Fig.~\ref{fig:fourth}(a), and are denoted by groups $\alpha$ and $\beta$. There, it can be seen that the variables in $X_{\textrm{Drug}}$ and $X_{\textrm{Meth}}$ show a contrasting pattern according to the two subgroups $\alpha$ and $\beta$.
Comparing $Z_{\textrm{Drug},S_2}$ (Fig.~\ref{fig:fourth}(a)) with the whole $X_{\textrm{Drug}}$ (Fig.~\ref{fig:fourth}(b)), we observe that the subgroups identified above are hidden in $X_{\textrm{Drug}}$.
Moreover, the subgroups $\alpha$ and $\beta$ are well-separated in the partially-joint score $\widehat{W}_{\{\textrm{Drug},{\textrm{Meth}}\}}$ of $Z_{\textrm{Drug},S_2}$ (Fig.~\ref{fig:fourth}(c)), while it is hard to find any subgroup in the principal component scores of whole $X_{\textrm{Drug}}$ (Fig.~\ref{fig:fourth}(d)).
The same conclusion can be made by inspecting the component of $X_{\textrm{Meth}}$ corresponding to the partially-joint score $\widehat{W}_{\{\textrm{Drug},{\textrm{Meth}}\}}$, denoted $Z_{\textrm{Meth},S_2}$, and the whole $X_{\textrm{meth}}$ in Fig.~\ref{fig:fourth}(e) and (f).
Thus we observed that PSI gives a more effective measure of finding inherent subgroups in a multi-omics data set than a separate application of the principal component analysis on each of the data matrix.

\begin{figure}[bt]
	\begin{center}
		\includegraphics[width=5in]{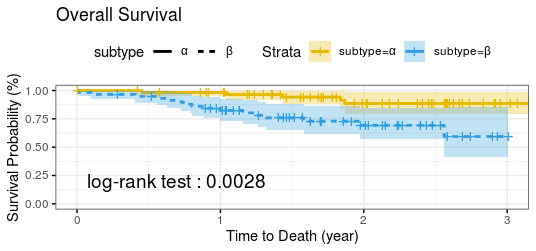}
	\end{center}
	\caption{ The difference in overall survival between the subgroups $\alpha$ and $\beta$ is displayed on a Kaplan plot with p-value 0.0028 from the log-rank test. \label{fig:fifth}}
\end{figure}

To find indicators that best explain the subgroups $\alpha$ and $\beta$, we conducted the Fisher exact test simultaneously on 59 gene mutations or chromosome defects of each patients, available as an ancillary information. The p-values from each Fisher exact test were adjusted by the Benjamini-Hochberg (BH) method. The smaller p-values indicate stronger differences of each mutation or chromosome defect between the subgroup $\alpha$ and $\beta$. We found that the immunoglobulin heavy chain variable (IGHV) region mutation status has the most associated relation with the subgroups at BH-adjusted p-value $1.026 \times 10^{-13}$. We also present the $2 \times 2$ table of the IGHV status and the subgroups, see Tables C.1 and C.2 in the supplementary material.
We postulate IGHV mutation status gives a substantial explanation for the subgrouping of CLL patients, that is, wild type matches to the subgroup $\alpha$ and mutation type to the subgroup $\beta$. Survival analyses on overall survival rates was also conducted, and we found statistically significant differences in survival between the two subgroups $\alpha$ and $\beta$ as shown in Fig.~\ref{fig:fifth}.

Again in Fig.~\ref{fig:fourth}(a), the variables in $Z_{\textrm{Drug},S_2}$ can be clustered into subgroups [a] and [b] showing a contrasting response pattern to the subgroups $\alpha$ and $\beta$ (variables showing weak responses were excluded as subgroup [c]). The subgroup [a] shows higher viability for $\beta$ (IGHV mutated) than $\alpha$ (IGHV wild type) and vice versa for [b]. Table C.3 (in the supplementary material) presents the list of prominent drugs that have appeared in subgroups [a] and [b] at least 4 times out of 5 concentrations. For the subgroup [a], the table lists a number of inhibitor drugs that target the B cell receptor (BCR) components, such as Bruton's tyrosine kinase (BTK; spebrutinib, ibrutinib), phosphatidylinositol 3-kinase (PI3K; idelalisib, duvelisib) and spleen tyrosine kinase (STK; tamatinib, PRT062607 HCL). AKT inhibitor(MK-2206) or SRC inhibitors (dasatinib) targets signal transduction pathways that promotes survival and growth of B cell lymphocytes. Unexpected encounter with HSP90 inhibitor (AT13387 or Onalespib) may be related to the stability of lymphocyte-specific SRC family kinases \citep{Mshaik2021}. The appearance of CHK inhibitors (PF 477736, AZD7762, CCT241533) may be linked to repairing mechanisms of DNA damages at G2 phase, known to be associated with WEE1 kinase and the AKT/PKB pathway \citep{Zhang2014}. For the subgroup [b], the appearance of mTOR inhibitor (everolimus) may suggest that mTOR pathway and shows different drug sensitivities to the BCR component, despite the fact that it is on the downstream of  AKR/PKB pathway. The role of IGHV in this implication requires further investigation. BCL2 inhibitor (navitoclax) and rotenone might be related to the role of mitochondria in apoptosis \citep{Wang2014}.

\section*{Supplementary Materials}

The supplementary materials contain the proofs of Lemma \ref{lem:lemone}, Theorems \ref{thm:thmone} and \ref{thm:thmtwo} and Corollary~\ref{cor:u}, a brief review on competing methods, detailed model settings used for simulations, additional numerical results and auxiliary information on real data analysis, including the table of drugs selected by the proposed method.

\bibliographystyle{elsarticle-harv}
\bibliography{Bibliography-MM-MC}





% \end{thebibliography}

\bigskip

\newpage

\bigskip

\appendix

\begin{center}
{\large\bf SUPPLEMENTARY MATERIAL}
\end{center}

\renewcommand\thesection{\Alph{section}}
\renewcommand{\thetable}{\Alph{section}.\arabic{table}}
\renewcommand{\thefigure}{\Alph{section}.\arabic{figure}}

\section{Proofs}

\subsection{Some basic facts for proof}

As it is the most elementary fact, we begin by recapitulating Lemma 2 and show its proof.

\begin{namedlemma}[2]
	For $i, j \in \mathcal{I}_K = \{ 1, \ldots, 2^K - 1\}$ and $S_i \cap S_j \neq \phi$, $[W_i] \perp [W_j]$.
\end{namedlemma}

\begin{proof}
	From the definition of the partially-joint score subspace, $[W_i]$, we immediately find that
	\begin{align*}
	[W_i] \perp [W_j], \quad i,j \in \mathcal{I}_K \;\; \textrm{and} \;\; S_j \cap S_i \neq \phi
	\end{align*}
	by the range-kernel complementarity property of the vector space projection transformation.
\end{proof}

We denote $\mathcal{I}_{<i} = \{ j : j < i,  S_j \cap S_i \neq \phi \}$. Hereafter $\mathcal{N}(T)$ is the null space of a linear transformation $T$ of $\mathbb{R}^n$ and $\mathcal{R}(T)$ its range space.

	\begin{lemma} \label{thm1.a}
		For $i \in \mathcal{I}_K$, we have $\mathcal{N} ( \bigcirc_{k \in \mathcal{I}_{<i}} P_k^{\perp}  ) = \oplus_{k \in \mathcal{I}_{<i}} [W_k]$.
	\end{lemma}
	
	\begin{proof}
		Let $v \in \oplus_{k \in \mathcal{I}_{<i}} [W_k]$. Then there exists a unique $\{ v_k \}_{k \in \mathcal{I}_{<i}}$ with $v_k \in [W_k]$ such that the sum of all $v_k$ is $v$.  For each $j_1 \in \mathcal{I}_{<i}$ and $v_{j_1} \in [W_{j_1}]$, it can be easily checked that $(\bigcirc_{k \in \mathcal{I}_{<j_1}} P_k^{\perp})(v_{j_1}) = \{0\}$ and then $(\bigcirc_{k \in \mathcal{I}_{<i}} P_k^{\perp})(v_{j_1}) = \{0\}$ follows. Thus $(\bigcirc_{k \in \mathcal{I}_{<i}} P_k^{\perp})(v) = \{ 0 \}$ and $\mathcal{N} ( \bigcirc_{k \in \mathcal{I}_{<i}} P_k^{\perp}  ) \supset \oplus_{k \in \mathcal{I}_{<i}} [W_k]$.
		
		Conversely, let $v' \not \in \oplus_{k \in \mathcal{I}_{<i}} [W_k]$. Then there exists a unique $\{ v_k \}_{k \in \mathcal{I}_{<i}}$ with $v_k \in [W_k]$ and non-zero $a \in (\oplus_{k \in \mathcal{I}_{<i}} [W_k])^\perp$ such that $v'$ is the sum of all $v_k$ and $a$. For each $j_1 \in \mathcal{I}_{<i}$ and $v_{j_1} \in [W_{j_1}]$, we have $(\bigcirc_{k \in \mathcal{I}_{<i}} P_k^{\perp})(v_{j_1}) = \{0\}$ as before. Since $a \perp [W_{k''}]$ for any $k'' \in \mathcal{I}_{<i}$, we also have $(\bigcirc_{k \in \mathcal{I}_{<i}} P_k^{\perp})(a) \neq 0$. Thus we have $(\bigcirc_{k \in \mathcal{I}_{<i}} P_k^{\perp})(v') \neq \{ 0 \}$ and $\mathcal{N} ( \bigcirc_{k \in \mathcal{I}_{<i}} P_k^{\perp}  ) \subset \oplus_{k \in \mathcal{I}_{<i}} [W_k]$.
	\end{proof}

	\begin{lemma} \label{thm1.b}
		For $i \in \mathcal{I}_K$, we have $\oplus_{k \in \mathcal{I}_{<i}} [W_k] = +_{k \in \mathcal{I}_{<i}} ( \cap_{k' \in S_k}  [V_{k'}] )$.
	\end{lemma}
	
	\begin{proof}
		We give a proof by induction on $k$. If $k = 1$, there is nothing to prove. If $k = 2$, $[W_1] = \cap_{k' \in S_1} [V_{k'}]$, so the statement is true for $k = 1, 2$.
		
		For any $k \geq 3$, suppose the statement holds, that is,
		\begin{align*}
			\oplus_{j \in \mathcal{I}_{< m}} [W_j] = +_{j \in \mathcal{I}_{<m}} ( \cap_{k' \in S_j}  [V_{k'}] )
		\end{align*}
		for all $1 \leq m \leq k$. Let $\tilde{k}$ be the largest element in $\mathcal{I}_{< k + 1}$. We denote $P = \bigcirc_{k' \in \mathcal{I}_{< \tilde{k}}} P_{k'}^{\perp}$ and $P^\perp$, the projection onto $\mathcal{N}(P)$ of $\mathbb{R}^n$. Then, we have
		\begin{align*}
			+_{j \in \mathcal{I}_{< k+1}} ( \cap_{k' \in S_j}  [V_{k'}] ) &= \oplus_{j \in \mathcal{I}_{<\tilde{k}}} [W_j] + \cap_{k' \in S_{\tilde{k}} }  [V_{k'}] \\
			&= \oplus_{j \in \mathcal{I}_{<\tilde{k}}} [W_j] + P(\cap_{k' \in S_{\tilde{k}} }  [V_{k'}]) + P^\perp(\cap_{k' \in S_{\tilde{k}} }  [V_{k'}])
		\end{align*}
		Indeed, $P^\perp(\cap_{k' \in S_{\tilde{k}} }  [V_{k'}]) \subset \mathcal{N}(P)$ and $\mathcal{N} (P) = \oplus _{k' \in \mathcal{I}_{<\tilde{k}}} [W_{k'}]$ by the previous lemma. Thus
		\begin{align*}
			+_{j \in \mathcal{I}_{< k+1}} ( \cap_{k' \in S_j}  [V_{k'}] ) &= \oplus_{j \in \mathcal{I}_{<\tilde{k}}} [W_j] + P(\cap_{k' \in S_{\tilde{k}} }  [V_{k'}]) \\
			&= \oplus_{j \in \mathcal{I}_{<\tilde{k}}} [W_j] \oplus [W_{\tilde{k}}] \\
			&= \oplus_{j \in \mathcal{I}_{<k+1}} [W_j].
		\end{align*}
		Therefore the statement holds for any $k \geq 3$ and the proof is completed.
	\end{proof}

%\subsection{Proof of Lemma 1}

	\begin{lemma} \label{lem:lemma1}
		For $k = 1,\ldots,K$,  $+_{i \in \{ k \in S_i\} } [W_{S_i}] = [V_k]$.
	\end{lemma}
	
\begin{proof}
    In Lemma A.2, we set $i$ such that $S_i = \{ k \}$. Then $\mathcal{I}_{<i} = \{ j : k \in S_j, |S_j| > 1 \}$ and $\oplus_{j \in \mathcal{I}_{<i}} [W_j] = +_{j \in \mathcal{I}_{<i}} ( \cap_{k' \in S_j}  [V_{k'}] )$. The proof goes similarly as in Lemma \ref{thm1.b}.  Let $\tilde{i}$ be the largest element in $\mathcal{I}_{< i}$. We denote $P = \bigcirc_{k' \in \mathcal{I}_{< \tilde{i}}} P_{k'}^{\perp}$ and $P^\perp$, the projection onto $\mathcal{N}(P)$ of $\mathbb{R}^n$.
    \begin{align*}
        +_{j \in \mathcal{I}_{< i} \cup S_i  } ( \cap_{k' \in S_j}  [V_{k'}] ) &= \oplus_{j \in \mathcal{I}_{<i}} [W_j] + \cap_{k' \in S_{i} }  [V_{k'}] \\
			&= \oplus_{j \in \mathcal{I}_{<i}} [W_j] + P(\cap_{k' \in S_{i} }  [V_{k'}]) + P^\perp(\cap_{k' \in S_{i} }  [V_{k'}]) \\
			&= \oplus_{j \in \mathcal{I}_{<i}} [W_j] + P(\cap_{k' \in S_{i} }  [V_{k'}]) \\
			&= \oplus_{j \in \mathcal{I}_{<i}} [W_j] \oplus [W_{i}] \\
			&= \oplus_{j \in  \mathcal{I}_{< i} \cup S_i} [W_j].
    \end{align*}
    By the set inclusion-exclusion principle, as $\mathcal{I}_{< i} \cup S_i$ involves every index-set that contains $k$, it is immediate that $+_{j \in \mathcal{I}_{< i} \cup S_i  } ( \cap_{k' \in S_j}  [V_{k'}] ) = [V_k]$.
\end{proof}

\subsection{Proof of Theorem 1}

\begin{proof}
	We claim that $\textrm{rank} ( [W_i] )$ is uniquely determined. By Sylvester's law of nullity, we have
	\begin{align*}
		\textrm{rank} ( [W_i] ) = \textrm{rank} ( \cap_{k \in S_i} [V_k] ) - \textrm{rank} (  \mathcal{N} (\bigcirc_{j \in \mathcal{I}_{<i}} P_j^{\perp}  )  \cap (\cap_{k \in S_i} [V_k])  ),
	\end{align*}
	where $\mathcal{I}_{<i} = \{ j : j < i,  S_j \cap S_i \neq \phi \}$ as before.
	By lemma \ref{thm1.a} and \ref{thm1.b}, Sylvester's law of nullity for our theorem is restated as
	\begin{align*}
		\textrm{rank} ( [W_i] ) = \textrm{rank} ( \cap_{k \in S_i} [V_k] ) - \textrm{rank} \left( ( +_{k \in \mathcal{I}_{<i}} ( \cap_{k' \in S_k}  [V_{k'}] ) )   \cap (\cap_{k \in S_i} [V_k])  \right).
	\end{align*}
	We want to make this expression in a more explicit form. For that, we suggest the following assertions.
	
	Let $l = |S _i|$. We set $[I_l] = +_{k \in \{ k : |S_k| > l \}} ( \cap_{k' \in S_k} [V_{k'}] )$. The projection onto $[I_l]$ in $\mathbb{R}^n$ is denoted by $P_{I_l}$ and the projection onto $[I_l]^\perp$ is denoted by $P_{I_l}^\perp$. Moreover, we consider an index set $ \mathcal{J}_{i, > l} =  \mathcal{I}_{<i} \cap \{ k: |S_k| > l \} $ and let $[J_i] = +_{k \in \mathcal{J}_{i, > l} } ( \cap_{k' \in S_k} [V_{k'}] )$. The projection onto $[J_i]$ in $\mathbb{R}^n$ is denoted by $P_{J_i}$ and the projection onto $[J_i]^\perp$ by $P_{J_i}^\perp$. Finally, we define $ \mathcal{J}_{i,  l} =  \mathcal{I}_{<i} \cap \{ k: |S_k| = l \}$.

	\begin{lemma} \label{thm1.c}
		$P_{I_l}^\perp = P_{I_l}^\perp \circ P_{J_{i}}^\perp$.
	\end{lemma}
	\begin{proof}
		Trivial from the fact $\mathcal{N} (P_{J_i}^\perp) \subset \mathcal{N} (P_{I_l}^\perp)$.
	\end{proof}
	
	\begin{lemma}\label{thm1.d}
		If $v_j \in P_{J_i}^\perp ( \cap_{k' \in S_j}  [V_{k'}] )$ with $j \in \mathcal{J}_{i,l} \setminus \mathcal{J}_{i,l}'$, then $v_j \in (\cap_{k' \in S_j}[V_{k'}]) \cap (\cap_{k' \in S_{m}}[V_{k'}])$ for some $m < i$, $m \neq j$ such that $S_{m} \cap S_i = \{ 0 \}$ and not for any $m < i$, $m \neq j$ such that $S_{m} \cap S_i \neq \{ 0 \}$.
	\end{lemma}
	\begin{proof}
		Consider the cases
		\begin{itemize}
			\item[(1)] $v_j \in (\cap_{k' \in S_j}[V_{k'}]) \cap (\cap_{k' \in S_{m}}[V_{k'}])$ for some $m < i$, $m \neq j$ such that $S_{m} \cap S_i \neq \{ 0 \}$ but not for any $S_{m} \cap S_i = \{ 0 \}$,
			\item[(2)] $(\cap_{k' \in S_j}[V_{k'}]) \cap (\cap_{k' \in S_{m}}[V_{k'}]) = \{ 0 \}$ for only $m \in \mathcal{J}_{i,l}$
		\end{itemize}
		
		In case (1), $v_j$ becomes automatically an element of $\cap_{k' \in S_t}[V_{k'}]$ such that $S_j \subset S_t$ and $S_m \subset S_t$. Then $v_j \in [I_i]$ since $S_i \cap S_t \neq \phi$ and $|S_t| > |S_i|$, and this is a contradiction. In case (2), $v_j \not\in \mathcal{R}(P_{I_l})$ and this is contradict to the assumption.
	\end{proof}

	\begin{proposition}\label{thm1.pro1}
		If $[V_k]_{k \in \mathcal{K}}$ is relatively independent, then $+_{k \in \mathcal{J}_{i,l} } P_{J_i}^\perp ( \cap_{k' \in S_k}  [V_{k'}] )$ and $P_{J_i}^\perp (\cap_{k \in S_i} [V_k])$ are independent.
	\end{proposition}
	
	\begin{proof}

		We want to show that the relative independence of $[V_k]_{k \in \mathcal{K}}$ is violated if $+_{k \in \mathcal{J}_{i,l} } P_{J_i}^\perp ( \cap_{k' \in S_k}  [V_{k'}] )$ and $P_{J_i}^\perp (\cap_{k \in S_i} [V_k])$ are linearly dependent. Suppose that there is nonzero $v \in +_{k \in \mathcal{J}_{i,l} } P_{J_i}^\perp ( \cap_{k' \in S_k}  [V_{k'}] ) \cap P_{J_i}^\perp (\cap_{k \in S_i} [V_k])$ with $l = |S_i|$. We are going to show that $+_{k \in \mathcal{J}_{i,l} } P_{J_i}^\perp ( \cap_{k' \in S_k}  [V_{k'}] ) \cap P_{J_i}^\perp (\cap_{k \in S_i} [V_k])$ is not $\{0\}$ in the following cases.
		
		First we exclude the case where $P_{I_l}^\perp(\cap_{k' \in S_i}  [V_{k'}] ) = \{ 0 \}$. If so we have $\cap_{k' \in S_i}  [V_{k'}] = \oplus_{k'' \in \{ k'' : S_i \subset S_{k''}  \} } [W_k''] = +_{k'' \in \{ k'' : S_i \subset S_{k''}  \} } [V_k'']$, then $\cap_{k' \in S_i}  [V_{k'}]$ itself is in $[I_i] = \mathcal{N} (P_{J_i}^\perp)$).
		
		Next, under the assumption that $P_{I_l}^\perp(\cap_{k' \in S_i}  [V_{k'}] ) \neq \{ 0 \}$, we run through the following situations. Now on $\mathcal{J}_{i,l}' = \{ k : |S_k| = l, k \in \mathcal{I}_{<i}, P_{I_i}^\perp(S_k) \neq \{ 0 \}    \}$.
		
		\begin{itemize}
			\item[(i)] $|\mathcal{J}_{i,l}| = 0$ : There is nothing to prove.
			
			\item[(ii)] $|\mathcal{J}_{i,l}| = 1$ and  $|\mathcal{J}_{i,l}'| = 0$ : \\ We will show that this case is vacuous. Suppose there exist nonzero $v \in P_{J_i}^\perp ( \cap_{k' \in S_i}  [V_{k'}] ) \cap P_{J_i}^\perp ( \cap_{k' \in S_k}  [V_{k'}] )$ with $k \in \mathcal{J}_{i,l}$. As $P_{I_l}^\perp ( \cap_{k' \in S_k}  [V_{k'}] ) = \{ 0 \}$, we can deduce that $P_{I_l}^\perp(v) = 0$.
			
			We pick a vector $u$ in $\cap_{k' \in S_j}  [V_{k'}]$ such that $S_i \subset S_j$ and $S_j \subset S_j$. As $\cap_{k' \in S_j}  [V_{k'}] \subset \cap_{k' \in S_i}  [V_{k'}]$ and $\cap_{k' \in S_j}  [V_{k'}] \subset \cap_{k' \in S_k}  [V_{k'}]$, we have $u \in (\cap_{k' \in S_i}  [V_{k'}]) \cap (\cap_{k' \in S_k}  [V_{k'}])$. As $S_i \cap S_j \neq \phi$, we observe that $\cap_{k' \in S_j}  [V_{k'}] \subset [I_i] = \mathcal{N} (P_{J_i}^\perp)$.
			\begin{align*}
				u &\in P_{J_i}( (\cap_{k' \in S_i}  [V_{k'}] ) \cap ( \cap_{k' \in S_k}  [V_{k'}])) \\ &\subset P_{J_i}(\cap_{k' \in S_i}  [V_{k'}]) \cap P_{J_i}(\cap_{k' \in S_k}  [V_{k'}]).
			\end{align*}
			
			Let $w = u + v$. Since $u \in P_{J_i}(\cap_{k' \in S_k}  [V_{k'}])$ and $v \in P_{J_i}^\perp(\cap_{k' \in S_k}  [V_{k'}])$ with respect to $S_k$ and the same for $S_i$, we find that $w \in (\cap_{k' \in S_i}  [V_{k'}]) \cap (\cap_{k' \in S_k}  [V_{k'}])$, and then, $w \in [I_i] = \mathcal{N}(P_{j_i}^\perp)$. But this forces $P_{J_i}^\perp(w) = v$ and $v$ to be zero and leads to vacuity.
			
			\item[(iii)] $|\mathcal{J}_{i,l}| = 1$ and  $|\mathcal{J}_{i,l}'| = 1$ : \\ Suppose there is nonzero $v \in P_{J_i}^\perp ( \cap_{k' \in S_k}  [V_{k'}] )$ with $k \in \mathcal{J}_{i,l}$. From Lemma \ref{thm1.c}, we have $P_{I_l}^\perp(v) \in P_{I_l}^\perp ( \cap_{k' \in S_k}  [V_{k'}] )$. By the same argument, $P_{I_l}^\perp(v) \in P_{I_l}^\perp ( \cap_{k' \in S_i}  [V_{k'}] )$. Then there is a nonzero $P_{I_l}^\perp(v) \in P_{I_l}^\perp ( \cap_{k' \in S_i}  [V_{k'}] ) \cap P_{I_l}^\perp ( \cap_{k' \in S_k}  [V_{k'}] )$.

			\item[(iv)] $|\mathcal{J}_{i,l}| \geq 2$, $|\mathcal{J}_{i,l}'| = 0$ and $v \in P_{J_i}^\perp ( \cap_{k' \in S_k}  [V_{k'}] )$ with some $k \in \mathcal{J}_{i,l}$ : the same as case (ii).

			\item[(v)] $|\mathcal{J}_{i,l}| \geq 2$, $|\mathcal{J}_{i,l}'| = 0$ and $v \not\in P_{J_i}^\perp ( \cap_{k' \in S_k}  [V_{k'}] )$ with any of $k \not\in \mathcal{J}_{i,l}$ : \\ We will show that this case is a generalization of case (ii) and also vacuous. As $v \in +_{k \in \mathcal{J}_{i,l}} P_{J_i}^\perp(\cap_{k' \in S_k}[V_{k'}])$, we express
			\begin{align*}
				v = \sum_{j \in \mathcal{J}_{i,l} } a_j v_j
			\end{align*}
			for each $a_j \in \mathbb{R}$ (at least two of them are nonzero) and $v_j \in P_{J_i}^\perp ( \cap_{k' \in S_j}  [V_{k'}] )$.

			By Lemma \ref{thm1.d}, for each $v_j$ for $j \in \mathcal{J}_{i,l}$, we can find $S_{m,j}$ such that $v_j \in S_{m,j}$ and $m < i$, $m \neq j$, $S_m \cap S_i = \{ 0 \}$. And this also implies that there exists certain $S_{t,j}$ and $v_j \in S_{t,j}$ such that $S_j \subset S_{t,j}$ and $S_{m,j} \subset S_{t,j}$. Here we point out that as $S_{t,j} \cap S_i \neq \phi$, since $S_{j} \cap S_i \neq \phi$. Then all $\cap_{k' \in S_{t,j}}  [V_{k'}]$ for $j \in \mathcal{J}_{i,l}$ and their linear combinations are subsets of $[I_i]$. This leads to the conclusion $v \in [I_i]$ and shows the vacuity of this case.

			\item[(vi)] $|\mathcal{J}_{i,l}| \geq 2$, $|\mathcal{J}_{i,l}'| \geq 1$ and $v \in P_{J_i}^\perp ( \cap_{k' \in S_k}  [V_{k'}] )$ with some $k \in \mathcal{J}_{i,l}$ :  \\ If $k \in \mathcal{J}_{i,l}'$, then the same as the case (iii). If $k \not\in \mathcal{J}_{i,l}'$, then the same as the case (ii).

			\item[(vii)] $|\mathcal{J}_{i,l}| \geq 2$, $|\mathcal{J}_{i,l}'| \geq 1$ and $v \not\in P_{J_i}^\perp ( \cap_{k' \in S_k}  [V_{k'}] )$ with any of $k \in \mathcal{J}_{i,l}$ :  \\
			As $v \in +_{k \in \mathcal{J}_{i,l}} P_{J_i}^\perp(\cap_{k' \in S_k}[V_{k'}])$, we express
			\begin{align*}
				v = \sum_{j \in \mathcal{J}_{i,l} } a_j v_j
			\end{align*}
			for each $a_j \in \mathbb{R}$ (at least two of them are nonzero) and $v_j \in P_{J_i}^\perp ( \cap_{k' \in S_j}  [V_{k'}] )$. Note that if $a_{j'} = 0$ for all $j' \in \mathcal{J}_{i,l}'$, then this case is essentially the same as case (v), so we only consider the situation at least one $a_{j'} \neq  0$ for $j' \in \mathcal{J}_{i,l}'$.
			
			By Lemma \ref{thm1.d} and its consequences in case (v), for each $j \in \mathcal{J}_{i,l} \setminus \mathcal{J}_{i,l}'$, there exists $S_{t,j}$ and $v_j \in S_{t,j}$ such that $S_j \subset S_{t,j}$ and $S_{m,j} \subset S_{t,j}$. As previously discussed, $\cap_{k' \in S_{t,j}}  [V_{k'}]$ for $j \in \mathcal{J}_{i,l} \setminus \mathcal{J}_{i,l}'$ is a subset of $[I_i]$. So we rule out the terms involving $j \in \mathcal{J}_{i,l} \setminus \mathcal{J}_{i,l}'$ and then
			\begin{align*}
				P_{I_l}^\perp (v) = \sum_{j' \in \mathcal{J}_{i,l}' } a_{j'} P_{I_l}^\perp(v_{j
					'}).
			\end{align*}
			Since each $\cap_{k' \in S_{j'}}  [V_{k'}] \neq \{ 0 \}$ for $j \in \mathcal{J}_{i,l}'$ is a subset of $[I_i]$ and at least one $a_{j'} \neq 0$, we deduce that $P_{I_l}^\perp (v)$ is non-zero. Therefore, the relative independence is violated.
			
		\end{itemize}
		
	\end{proof}

	Now the last term in the RHS of our law of nullity is re-expressed as
	\begin{align} \label{thm1_exp3}
		& ( +_{k \in \mathcal{I}_{<i}} ( \cap_{k' \in S_k}  [V_{k'}] ) )   \cap (\cap_{k \in S_i} [V_k]) \\
		= &  ( ( +_{k \in \mathcal{J}_{i, <l}} ( \cap_{k' \in S_k}  [V_{k'}] )) \oplus (+_{k \in \mathcal{J}_{i,l} } P_{J_i}^\perp ( \cap_{k' \in S_k}  [V_{k'}] ) ) ) \cap (\cap_{k \in S_i} [V_k]) \nonumber \\
		= &  ( ( +_{k \in \mathcal{J}_{i, <l}} ( \cap_{k' \in S_k}  [V_{k'}] )) \oplus (+_{k \in \mathcal{J}_{i,l} } P_{J_i}^\perp ( \cap_{k' \in S_k}  [V_{k'}] ) ) ) \nonumber \\ & \quad \cap ( P_{J_i} (\cap_{k \in S_i} [V_k]) \oplus P_{J_i}^\perp (\cap_{k \in S_i} [V_k])  )  \nonumber
	\end{align}
	In Proposition \ref{thm1.pro1}, we have observed that $+_{k \in \mathcal{J}_{i,l} } P_{J_i}^\perp ( \cap_{k' \in S_k}  [V_{k'}] )$ and $P_{J_i}^\perp (\cap_{k \in S_i} [V_k])$ are independent. As $P_{J_i} (\cap_{k \in S_i} [V_k]) \subset  +_{k \in \mathcal{J}_{i, <l}} ( \cap_{k' \in S_k}  [V_{k'}]$, the term (\ref{thm1_exp3}) becomes
	\begin{align*}
		( +_{k \in \mathcal{I}_{<i}} ( \cap_{k' \in S_k}  [V_{k'}] ) )   \cap (\cap_{k \in S_i} [V_k])
		=     ( +_{k \in \mathcal{J}_{i, <l}} ( \cap_{k' \in S_k}  [V_{k'}] ) \cap  (P_{J_i} (\cap_{k \in S_i} [V_k]) ).
	\end{align*}
	Finally, we have demonstrated
	\begin{align*}
		\textrm{rank} ( [W_i] ) &= \textrm{rank} ( \cap_{k \in S_i} [V_k] ) - \textrm{rank} \left( ( +_{k \in \mathcal{I}'_{< i} } ( \cap_{k' \in S_k}  [V_{k'}] ) )   \cap (\cap_{k \in S_i} [V_k])  \right) \\
		&= \textrm{rank} ( \cap_{k \in S_i} [V_k] ) \\
		& \;\;  - \textrm{rank} \left(  ( +_{k \in \mathcal{J}_{i, <l}} ( \cap_{k' \in S_k}  [V_{k'}] ) \cap  (P_{J_i} (\cap_{k \in S_i} [V_k]) ) \cap (\cap_{k \in S_i} [V_k]) \right) \\
		&= \textrm{rank} ( \cap_{k \in S_i} [V_k] ) - \textrm{rank} \left( P_{J_i} (\cap_{k \in S_i} [V_k]) \right).
	\end{align*}
	It is notable that the determination of $\textrm{rank}([W_i])$ depends only on $\mathcal{J}_{i, >l}$, that is, the set of indices $j$ such that $|S_j| > |S_i|$ and $S_j \cap S_i \neq \phi$. In other words, it does not depend on any index-sets of the same size as $S_i$ and their orderings.
	
	As partially-joint score subspaces $[W_i]$ are constructed recursively and the determination of each $[W_i]$'s rank only depends on $\mathcal{J}_{i,>l}$, not on the ordering index-sets of size $l = |S_i|$, we conclude that the set of pairs $\{ (S_i, r(S_i)) : i \in \mathcal{I}_K \}$ is unique.
	
\end{proof}

\subsection{Proof of Theorem 2}

\begin{proof}
	We give a proof recursively on $l \in \mathcal{K}$. If $l = K$, there only exists $[W_1] = \cap_{k' \in S_1} [V_{k'}]$ for $S_1 = \mathcal{K}$, so there is nothing to prove.  If $l = K - 1$, by absolute orthogonality, all $P_1^\perp(\cap_{k' \in S_{i} } [V_{k'}])$ are orthogonal each other for $i = \{ 2, \ldots, K+1 \}$. Therefore, in determining each $[W_i]$ for $i = \{ 2, \ldots, K+1 \}$, other $[W_t]$ for $t \in \mathcal{J}_{i,2} = \mathcal{I}_{<i} \cap \{i' : |S_{i'}| = 2 \}$ does not affect on the construction of $[W_i]$.

	For any $l \leq K - 2$, suppose the statement holds, that is, a partially-joint score subspace $[W_{j}]$ such that $|S_j| = l' > l$ is uniquely determined only by $[W_{j'}]$ for $S_{j'} > S_j$. For each index $i \in  \mathcal{J}_{l} = \{i' :|S_{i}| = l \}$ (regardless of ordering), we have $P_{I_l}(\cap_{k' \in S_{i}} [V_{k'}]) = P_{J_i}(\cap_{k' \in S_{i}} [V_{k'}])$ by absolute orthogonality.  Then among all indices $i' \in \mathcal{J}_{l}$, $P_{J_{i'}}^\perp(\cap_{k' \in S_{i'} } [V_{k'}])$ are orthogonal each other.
	
	Suppose an ordering on the set of all index-sets of size $l$ is given, denoted by $( S_{i_1}, \ldots, S_{i_h}  )$ with $h = {}_KC_l$. For $i_1$, $[W_{i_1}]$ is just determined as $P_{J_{i_1}}^\perp(\cap_{k' \in S_{i_1} } [V_{k'}])$. Next for $i_2$, as $[W_{i_1}]$ and  $P_{J_{i_2}}^\perp(\cap_{k' \in S_{i_2} } [V_{k'}])$ are orthogonal, we check that $P_{i_1}^\perp \circ P_{J_{i_2}}^\perp(\cap_{k' \in S_{i_2} } [V_{k'}]) = P_{J_{i_2}}^\perp(\cap_{k' \in S_{i_2} } [V_{k'}])$. Thus $[W_{i_2}]$ is determined regardless of $[W_{i_1}]$. In recursive manner, for $i \in \{ i_3, \ldots, i_h \}$, $[W_{i}]$ is determined regardless of all $[W_{i'}]$ for $i' \in \mathcal{J}_{i,l}$, or in other word, is uniquely determined as $P_{J_{i}}^\perp(\cap_{k' \in S_{i} } [V_{k'}])$ depending only on $[W_{j'}]$s for $S_{j'} > l$.
	
\end{proof}

\subsection{Proof of Corollary 1}

\begin{proof}
	From the discussions of Section 2, we know that $[W_{(k.)}]$ is indeed the score subspace of $Z_k$ for each $k = 1, \ldots, K$. Thus, given unique $[W_i]$'s for $i \in \mathcal{I}_{(k)}$ from Theorem 2, the subspace $[U_{(k),i}]$ generated by $U_{(k),i}$ is unique.
\end{proof}

\subsection{Examples}

The following two examples presents the cases where relative independence is satisfied and not:

\begin{example} \label{uniqueex1}
	Let $K = 3$, $n = 4$ and
	\begin{align*}
	V_1 = \begin{pmatrix} 1 & 0 \\  0 & 1 \\ 0 & 0 \\ 0 & 0 \end{pmatrix}, \quad V_2 = \begin{pmatrix} 1 & 0 \\ 0 & 1 / \sqrt{2} \\ 0 & 1 / \sqrt{2} \\ 0  & 0 \end{pmatrix}, \quad V_3 = \begin{pmatrix} 1 & 0 \\ 0 & 0 \\ 0 & 1 / \sqrt{2} \\ 0 & 1 / \sqrt{2}  \end{pmatrix}
	\end{align*}
	then $[I_1] = [(1,0,0,0)^T]$ and $P_{I_1}^\perp [V_1] = [(0,1,0,0)^T]$, $P_{I_1}^\perp [V_2] = [(0,1/\sqrt{2},1/\sqrt{2},0)^T]$ and $P_{I_1}^\perp [V_3] = [(0,0,1/\sqrt{2},1/\sqrt{2})^T]$ are linearly independent. Thus $\{ [V_1], [V_2], [V_3]  \}$ is relatively independent.
\end{example}

\begin{example} \label{uniqueex2}
	Let $K = 3$, $n = 4$ and
	\begin{align*}
	V_1 = \begin{pmatrix} 1 & 0 \\  0 & 1 \\ 0 & 0 \\ 0 & 0 \end{pmatrix}, \quad V_2 = \begin{pmatrix} 1 & 0 \\ 0 & 1 / \sqrt{2} \\ 0 & 1 / \sqrt{2} \\ 0  & 0 \end{pmatrix}, \quad V_3 = \begin{pmatrix} 1 & 0 & 0\\ 0 & 0 & 0\\ 0 & 1 & 1 / \sqrt{2} \\ 0 & 0 & 1 / \sqrt{2}  \end{pmatrix}
	\end{align*}
	As $P_{I_1}^\perp [V_3] \cap (P_{I_1}^\perp [V_1] + P_{I_1}^\perp [V_2] ) = [(0,0,1,0)^T]$ is not empty, thus $\{ [V_1], [V_2], [V_3]  \}$ is not relatively independent.
\end{example}

We present the following examples that support Theorem 1, that relative independence indeed guarantee the uniqueness of the \textit{partially-joint structure} $\mathfrak{S} = \{ (S_i, r(S_i)) : i \in \mathcal{I}_K, r(S_i) > 0  \}$.

\begin{example}[cont'd from Example \ref{uniqueex1}.]
	Under the ordering $S_1 = \{ 1,2,3 \}, S_5 = \{ 1 \}, S_6 = \{2 \}, S_7 = \{ 3 \}$, we obtain $[W_1] = [(1,0,0,0)^T]$, $[W_5] = [(0,1,0,0)^T]$, $[W_6] = [(0,0,1,0)^T]$ and $[W_7] = [(0,0,0,1)^T]$. Then we have
	\begin{align*}
	\mathfrak{S} = \{ (\{1,2,3\},1), (\{1\},1),(\{2\},1),(\{3\},1)  \}. \end{align*}
	On the other hand, in the case $S_1 = \{ 1,2,3 \}, S_5 = \{ 2 \}, S_6 = \{1 \}, S_7 = \{ 3 \}$, we obtain $[W_1] = [(1,0,0,0)^T]$, $[W_5] = [(0,1 / \sqrt{2},1 / \sqrt{2},0)^T]$, $[W_6] = [(0,-1 / \sqrt{2},1 / \sqrt{2},0)^T]$ and $[W_7] = [(0,0,0,1)^T]$. The partially-joint structure is still the same as above.
\end{example}

\begin{example}[cont'd from Example \ref{uniqueex2}.]
	Under the ordering $S_1 = \{ 1,2,3 \}, S_5 = \{ 1 \}, S_6 = \{2 \}, S_7 = \{ 3 \}$, we obtain $[W_1] = [(1,0,0,0)^T]$, $[W_5] = [(0,1,0,0)^T]$, $[W_6] = [(0,0,1,0)^T]$ and $[W_7] = [(0,0,0,1)^T]$. The partially-joint structure is \begin{align*} \mathfrak{S} = \{ (\{1,2,3\},1), (\{1\},1),(\{2\},1),(\{3\},1)  \}. \end{align*}
	On the other hand, in the case $S_1 = \{ 1,2,3 \}, S_5 = \{ 3 \}, S_6 = \{2 \}, S_7 = \{ 1 \}$, we obtain $[W_1] = [(1,0,0,0)^T]$, $[W_5] = [(0,1,0,0)^T, (0,0,1 / \sqrt{2},1 / \sqrt{2})^T]$, $[W_6] = [(0,0,1 / \sqrt{2},-1 / \sqrt{2})^T]$ and $[W_7] = \{0\}$. This time, the partially-joint structure is \begin{align*} \mathfrak{S} = \{ (\{1,2,3\},1), (\{1\},0),(\{2\},1),(\{3\},2)  \}. \end{align*}
\end{example}

We also present the following examples that support Theorem 2, that absolute orthogonality guarantee the uniqueness of the partially-joint score subspaces.

\begin{example}
	Let $K = 4$, $n = 7$ and
	\begin{align*}
	V_1 &= \begin{pmatrix}   0 & 0 & 1 & 0 & 0 & 0 \end{pmatrix}^T \\
	V_2 &= \begin{pmatrix}  0 & 0 & 0 & 1 & 0  & 0 \end{pmatrix}^T \\
	V_3 &= \begin{pmatrix}  1/\sqrt{2} & 1/\sqrt{2} & 0 & 0 & 0  & 0 \\  0 & 0 & 0 & 0 & 1 & 0 \end{pmatrix}^T \\
	V_4 &= \begin{pmatrix}   1/\sqrt{2} & 1/\sqrt{2} & 0 & 0 & 0  & 0 \\  0 & 0 & 0 & 0 & 0 & 1 \end{pmatrix}^T.
	\end{align*}
	This example satisfies absolute orthogonality. Between under two orderings $(S_{11} = \{3,4\}, S_{12} = \{1\}, S_{13} = \{2\})$ and $(S_{11} = \{3,4\}, S_{12} = \{2\}, S_{13} = \{1\})$, the determinations of $[W_{12}]$ and $[W_{13}]$ are the same.
\end{example}

\begin{example}
	Let $K = 4$, $n = 7$ and
	\begin{align*}
	V_1 &= \begin{pmatrix}   1/2\sqrt{2} & \sqrt{3}/2\sqrt{2} & 1/\sqrt{2}& 0 & 0 & 0 \end{pmatrix}^T \\
	V_2 &= \begin{pmatrix}  \sqrt{3}/2\sqrt{2} & 1/2\sqrt{2} & 0 &1/\sqrt{2}& 0  & 0 \end{pmatrix}^T \\
	V_3 &= \begin{pmatrix}  1/\sqrt{2} & 1/\sqrt{2} & 0 & 0 & 0  & 0 \\  0 & 0 & 0 & 0 & 1 & 0 \end{pmatrix}^T \\
	V_4 &= \begin{pmatrix}   1/\sqrt{2} & 1/\sqrt{2} & 0 & 0 & 0  & 0 \\  0 & 0 & 0 & 0 & 0 & 1 \end{pmatrix}^T.
	\end{align*}
	This example satisfies relative orthogonality, but not absolute orthogonality. Between under two orderings $(S_{11} = \{3,4\}, S_{12} = \{1\}, S_{13} = \{2\})$ and $(S_{11} = \{3,4\}, S_{12} = \{2\}, S_{13} = \{1\})$, the determinations of $[W_{12}]$ and $[W_{13}]$ are not the same because $[V_1]$ and $[V_2]$ are not orthogonal and $[W_{11}] = (1/\sqrt{2}, 1/\sqrt{2},0,0,0,0)^T$ does not have an effect on the determination of $[W_{12}]$ and $[W_{13}]$ by definition.
\end{example}

\section{Additional Information on Simulation Study}

\subsection{Example of the Measure of Dissimilarity between Two Partially-Joint Structures}

For the comparison between two partially-joint structure, we devised a following concept of the dissimilarity measure between two partially-joint structures.

We turn a partially-joint structure into the corresponding binary structure matrix. For example, let $K = 3$ and partially-joint structure be
\begin{align*}
\mathfrak{S} = \{  ( \{1,2,3\} , 2 , W_1) , ( \{1,2\}, 1, W_2 ) , ( \{1,3\}, 1,W_3 ) , ( \{2,3\}, 1 , W_4) , ( \{3\}, 1, W_7 ) \}.
\end{align*}
Then we have the corresponding binary structure matrix
\begin{align*}
T = \begin{pmatrix} 1 & 1 & 1 & 1 & 0 & 0 \\ 1 & 1 & 1 & 0 & 1 & 0 \\ 1 & 1 & 0 & 1 & 1 & 1 \end{pmatrix}.
\end{align*}

Consider two binary structure matrices $T_1 \in \{ 0 ,1 \}^{K \times m_1}$ and $T_2 \in \{ 0, 1 \}^{K \times m_2}$. Discarding all the identical columns, we denote the remaining columns $\widetilde{T}_1$ and $\widetilde{T}_2$. For example, from
\begin{align*}
T_1 = \begin{bmatrix}1 & 1 \\ 1 & 1 \\ 1 & 0 \end{bmatrix}, \qquad  T_2 = \begin{bmatrix}1 & 1 & 0 \\ 1 & 1 & 1\\ 1 & 1 & 1 \end{bmatrix},
\end{align*}
we obtain
\begin{align*}
\widetilde{T}_1 =  \begin{bmatrix}1  \\  1 \\  0 \end{bmatrix}, \qquad  \widetilde{T}_2 = \begin{bmatrix}1 & 0  \\  1 & 1\\ 1 & 1 \end{bmatrix}.
\end{align*}

For each remaining column of $\widehat{T}_1$ (or $\widehat{T}_2$), find the closest column of $\widehat{T}_2$ (or $\widehat{T}_1$) in the Hamming distance. The dissimilarity measure between $\widehat{T}_1$ and $\widehat{T}_2$ is the sum of the squares of all these Hamming distance.

In the example above, for $(1,1,0)^T$ of $\widetilde{T}_1$, the closest column of $\widetilde{T}_2$ is $(1,1,1)^T$ and the difference is $1$. For $(1,1,1)^T$ of $\tilde{T}_2$, the difference between $(1,1,0)^T$ is $1$ and for $(0,1,1)^T$ of $\tilde{T}_2$, it is $2$. The dissimilarity measure is then $1^2 + 1^2 + 2^2 = 6$.

\subsection{Review on Methodology of Other Methods}

We briefly review the methodology of AJIVE \citepsm{Feng2018}, SLIDE \citepsm{Gaynanova2019}, COBS \citepsm{Gao2020} and JIVE \citepsm{Lock2013}.

% AJIVE

\textbf{AJIVE} In AJIVE, each signal matrix $Z_k \in \mathbb{R}^{p_k \times n}$ is regarded as a sum of joint structure $J_k$ and individual structure $I_k$ for $k = 1, \ldots, K$. A joint structure $J_k$ is viewed as the score subspace $[V_M] \in \mathbb{R}^n$ shared by all $Z_i$'s.

AJIVE extracts each estimated signal matrix $\widehat{Z}_k$ from dataset $X_k$ and $\widehat{Z}_k$ is of initial rank estimate $\widetilde{r}_k$. The estimated shared joint component $[\widehat{V}_M]$ is obtained as a flag mean among score subspaces of $\widehat{Z}_k$'s, or $[\widehat{V}_1], \ldots, [\widehat{V}_k]$, in a sense of the projection Frobenius norm distance as our method. The rank $r_J$ of $[\widehat{V}_M]$ (called the joint rank) is estimated using the simulated distribution of the largest singular value of the concatenated matrix of random directions and that of Wedin bounds:
\begin{enumerate}
	\item[(1)] If the largest squared singular value of the column concatenation matrix $\widehat{V}$ of $\widehat{V}_k$'s is larger than the $5$th percentile of the simulated distribution of the largest squared singular value of the concatenation matrix of random orthogonal matrices of the same size as $\widehat{V}_k$'s (or random direction bound), then $[\widehat{V}_M]$ is not generated by noise in $95$ percent of probability.
	\item[(2)] If there are $\widehat{r}_J$ squared singular values of $\widehat{V}$ are larger than the $95$th percentile of the simulated distribution of Wedin bounds, then the first $\widehat{r}_J$ right singular vectors are used as the basis for the estimated joint score subspace $[\widehat{V}_M]$.
\end{enumerate}
The estimated joint structure $\widehat{J}_k$ is a projection of the dataset $X_k$ onto the estimated joint score subspace $[\widehat{V}_M]$, that is, $\widehat{J}_k = X_k \widehat{V}_M \widehat{V}_M^T$. Each estimated individual structure $\widehat{I}_k$ is obtained as $X_k \cdot (I - \widehat{V}_M \widehat{V}_M^T)$. The row spaces of each estimated individual structure $\widehat{I}_k$ is orthogonal to $[\widehat{V}_M]$. There is no guarantee that individual structures are mutually orthogonal. The joint score matrix is just defined as $\widehat{V}_M$ and the corresponding joint loading matrix for $k$th data source is a regression of $\widehat{J}_k$ on $\widehat{V}_M^T$, computed as $\widehat{J}_k \cdot \widehat{V}_M$. The individual loading and score matrices are obtained from SVDs of $\widehat{I}_k$'s.

% SLIDE

\textbf{SLIDE} SLIDE identifies the partially-joint structure with the penalized matrix factorization, that is,
\begin{align*}
(\widetilde{U}, \widetilde{V})= \underset{U, V}{\arg\min} \sum_{k = 1}^K \frac{1}{2} \Vert X_k - U_k V^T \Vert_F^2 + \lambda \sum_{j = 1}^r \Vert U_{kj} \Vert_2 \quad \textrm{s.t.} \;\; V^T V = I,
\end{align*}
where $U_{kj}$ is the $j$th column of the loading matrix $U_k$ of $X_k$ and $r$ is the number of all possible sparsity patterns. After computing $\widetilde{U}$ and $\widetilde{V}$ with an iterative algorithm, the corresponding structure matrix $\widehat{T}$ is obtained from the sparse structure of $\widetilde{U}$. Note that the concept of the structure matrix $\widehat{T}$ here is identical to the partially-joint structure matrix of ours in Section B.1. Even though this optimization problem is nonconvex and there is no guarantee about convergence to the global optimum, authors reported that a local solution can be obtained heuristically by initializaing $V$ with the left singular matrix of concatenated $X$ that works well in the simulation.

Then SLIDE estimate the loading and score matrices, $\widehat{U}$ and $\widehat{V}$, for the structure $\widehat{T}$ by solving the following optimization problem with an iterative algorithm,
\begin{align*}
(\widehat{U}, \widehat{V})= \underset{U, V}{\arg\min} \Vert X - UV^T \Vert_F^2 \quad \textrm{s.t.} \;\; V^T V = I,
\end{align*}
with the constraint that the loading $U$ has the same sparsity structure as $\widehat{T}$.

In model validation, SLIDE adapt the block cross validation (BCV) procedure to select the best structure $\widehat{T}_{\textrm{best}}$. BCV splits rows and columns of each dataset $X_k$ into submatrices $X_{k}^{11}, X_{k}^{12}, \ldots, X_{k}^{21}, \ldots$. Then it holds out a set of submatrices $X^{ij} = [ X_{1}^{ij}, X_{2}^{ij}, \ldots, X_{K}^{ij} ]$ of the same sub-block position in each dataset $X_k$ and evaluate the prediction error on $X^{ij}$'s. Given a set of structure candidates $\widehat{T}_1, \widehat{T}_2, \ldots$, we select one that minimizes the error across all folds.

% JIVE

\textbf{JIVE} Like in AJIVE, JIVE decompose each signal matrix $X_i$ as a sum of joint structure $J_i$ and individual structure $I_i$, for $i = 1, \ldots, K$. After defining $R$ to be a row concatenation of $R_i = X_i - J_i - I_i$, JIVE estimate both joint and individual structures by minimizing $\Vert R \Vert_F^2$ under the given ranks. An alternating iterative algorithm is implemented for the estimation finding individual component with given joint component at one step and vice versa at another step. The estimated joint structure is identical to the first $r_J$ terms in the SVD of $X$ with individual components removed and the estimated individual structures to the first $r_i$ terms in the SVD of $X_i$ with the joint component removed. The selection of $r_J$ and $r_i$s are validated using the permutation test.

% COBS

\textbf{COBS} COBS iteratively estimates a sequence of loading vectors, $u_i$ for $i = 1, \ldots, r$ for given $r$, while updating the data matrix $X = [X_1, \ldots, X_k] \in \mathbb{R}^{n \times \sum p_k}$. The algorithm starts with $X^{[0]} = X$. At the $i$th step, with the current data matrix $X^{[i-1]}$, the $i$th loading vector $u_i$ is estimated solving the following maximization problem, that is,
\begin{align*}
\widetilde{u}_i =  \underset{u}{\max} \Vert (X^{[i - 1]})^T u \Vert_2^2 \quad \textrm{s.t.} \;\; u^T u = 1.
\end{align*}
As each $u_i$ is equipped with block structure $u_i = \left(u_{(1)}^T \ldots u_{(K)}^T \right)^T$, we can give sparsity at two levels of thresholding, one for block-wise sparsity and the other for overall sparsity in estimating $\widetilde{u}_i$. The tuning parameters $\alpha_v \in [0,1]$ and $\lambda_v \geq 0$ control the two levels of sparsity respectively, and $\widehat{u}$ is thresholded as a normalized solution of
\begin{align*}
\min_x \frac{1}{2} \Vert x - \widetilde{u}_i \Vert^2 + \gamma_1 \Vert x \Vert_1 + \gamma_2 \Vert x \Vert_{2,1},
\end{align*}
where $\gamma_1 = \alpha_v \lambda_v$ and $\gamma_2 = (1 - \alpha_v) \lambda_v$. Then the score vector $v_i$ is estimated as the empirical BLUP and dataset $X^{[i-1]}$ is updated as $X^{[i]} = X^{[i-1]} - \widehat{u}_i \widehat{v}^T$.

\clearpage

\subsection{Computation time comparison}

%% Time comparison

\begin{table}[hp]
		\caption{Time comparison among our method and other four method(SLIDE, COBS, AJIVE and JIVE). The computation time reported. The unit is second(s). Averages and standard deviations are over 10 replications} \label{Time.comp}
	\centering
	\resizebox{\textwidth}{!}{%
		\begin{tabular}{ccccccc}
			\hline
			Model & SNR & PSI & SLIDE bcr & COBS & AJIVE & JIVE \\
			\hline
			1 & 15 & \textbf{0.78 (0.1)} & 283.61 (8.37) & 11.36 (0.19) & 12.01 (0.13) & 6.32 (0.15) \\
 & 10 & \textbf{0.77 (0.06)} & 281.73 (5.77) & 10.9 (0.13) & 12.06 (0.14) & 148.55 (74.98) \\
   & 5 & \textbf{0.76 (0.02)} & 362.66 (3.67) & 11.33 (0.09) & 12.01 (0.1) & 108.64 (0.57) \\
			\hline
			
2 & 15 & \textbf{0.95 (0.15)} & 283.91 (6.74) & 3.85 (0.15) & 12.35 (0.36) & 19.83 (0.5) \\
   & 10 & \textbf{0.86 (0.03)} & 332.45 (6.2) & 3.54 (0.15) & 12.04 (0.14) & 24.44 (0.25) \\
   & 5 & \textbf{0.87 (0.08)} & 361.34 (9.2) & 3.09 (0.07) & 12.1 (0.06) & 51.79 (1.21) \\
			\hline
			
			3 & 15 & \textbf{1.73 (0.13)} & 265.59 (5.26) & 11.83 (0.48) & 11.92 (0.16) & 56.79 (1.09) \\
   & 10 & \textbf{1.62 (0.09)} & 287.99 (5.2) & 10.07 (0.18) & 12.01 (0.41) & 148.38 (3.41) \\
   & 5 & \textbf{1.52 (0.08)} & 398.44 (6.75) & 11.6 (0.19) & 12.06 (0.27) & 65.95 (1.82) \\
			\hline
			
			4 & 15 & \textbf{1.48 (0.12)} & 237.01 (4.61) & 14.52 (0.36) & 12.57 (0.97) & 67.55 (1.47) \\
 & 10 & \textbf{1.39 (0.04)} & 280.57 (4.04) & 16.39 (0.15) & 12.04 (0.12) & 28.72 (0.52) \\
   & 5 & \textbf{1.41 (0.09)} & 360.8 (4.95) & 15.02 (0.12) & 12 (0.09) & 37.87 (0.16) \\
			\hline

		5 & 15 & \textbf{2.45 (0.13)} & 240.77 (5.24) & 15.65 (0.18) & 12.01 (0.19) & 100.54 (2.14) \\
   & 10 & \textbf{2.36 (0.06)} & 252.43 (5.94) & 16.2 (0.2) & 11.95 (0.12) & 70.98 (8.43) \\
   & 5 & \textbf{2.36 (0.11)} & 359.94 (5.37) & 15.47 (0.14) & 11.92 (0.12) & 108.68 (15.49) \\
			\hline
			
		6 & 15 & \textbf{9.07 (0.68)} & 599.56 (8.64) & 70.99 (3.57) & 32.79 (1.91) & 439.35 (16.24) \\
   & 10 & \textbf{7.86 (0.96)} & 770.04 (13.9) & 72.64 (2.43) & 32.77 (1.99) & 297.46 (9.04) \\
   & 5 & \textbf{8.84 (0.61)} & 940.57 (37.64) & 73.77 (3.87) & 32.79 (2.24) & 417.45 (63.1) \\
			\hline
	\end{tabular} }
\end{table}

\subsection{Results on Tuning Parameter Selection} \label{sec:sec5.2}

We report the performances of our tuning parameter selection procedure of Section 3.3 using the six models given in Section 5. %, as discussed in Section \ref{sec:exmample}.
Since the tuning parameter $\lambda$ represents a threshold for principal angles, the candidates for $\lambda$ are given by $\lambda = 0^\circ, 1^\circ, \ldots, 90^\circ$. For each value of $\lambda$, we evaluated the empirical risk.
As discussed in Section 3.3, we take the parameter $\tilde{\lambda}$ that gives the smallest empirical risk and also compare
$d( \widehat{\mathfrak{S}}_{tr}(\widetilde{\lambda}), \widehat{\mathfrak{S}}(\lambda)$ as a function of $\lambda$.
We also present $d (\widehat{\mathfrak{S}} (\lambda), \mathfrak{S} )$, which reflects how much the estimated structure differs from the true structure on each value of angle threshold, under the  situation where the true structure (`oracle') is known.

\begin{figure}[bt]
	\begin{center}
		\includegraphics[width=1\textwidth]{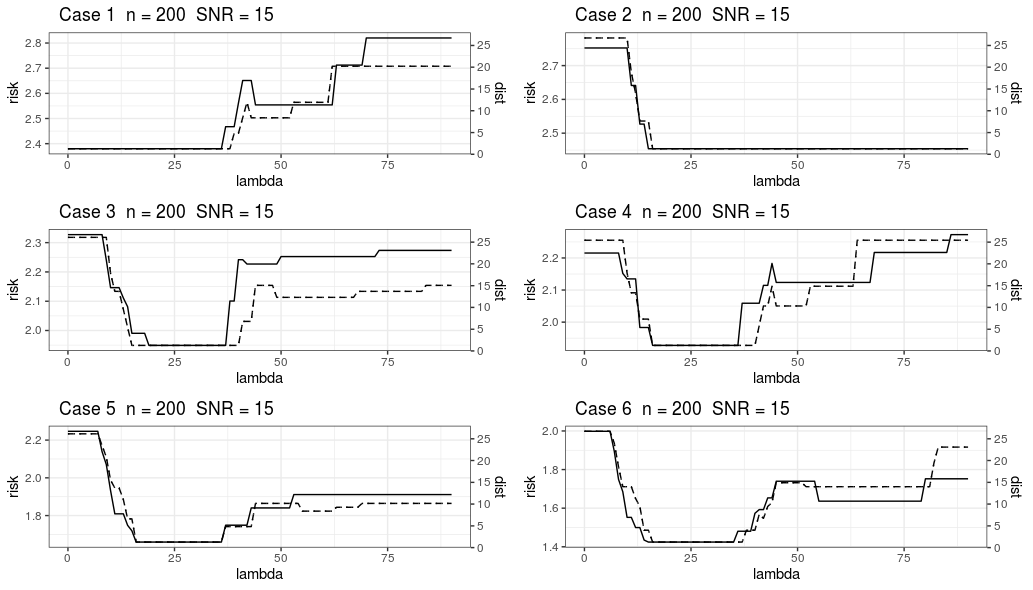}
	\end{center}
	\caption{The values of empirical risk (solid) and $d( \widehat{\mathfrak{S}}_{tr}(\widetilde{\lambda}), \widehat{\mathfrak{S}}(\lambda)$ (dotted), $d (\widehat{\mathfrak{S}} (\lambda), \mathfrak{S} )$, (dashed).
\label{fig:second}}
\end{figure}

First, consider $\textrm{SNR} = 10$. The empirical risk is minimized at an interval of $\lambda$'s, and for any $\lambda$ in the interval, the corresponding structure $\widehat{\mathfrak{S}}$ matches the true $\mathfrak{S}$; the valley bottoms of empirical risk (solid line) are posited inside those of the measure of dissimilarity (dashed line) with value zero, as seen in Fig. \ref{fig:second}. For each model in the figure, solid line (empirical risk) shows a similar shape as dashed line (the measure of dissimilarity), which implies that empirical risk well reflects the difference between the estimated structure and the true structure. %In Model 6, the true signal rank is $8$ for each dataset, for instance, from $\{ (\{1,2,3\},2),(\{1,2\},2),(\{1,3\},2),(\{1\},2)\}$ in the case $X_1$. However, \texttt{getnfrac} function only estimated $r_1 = 8$, $r_2 = 8$ and $r_3 = 7$. The estimated partially-joint structure lacks $(\{3\},1)$ from the true structure.

\begin{figure}[bt]
	\begin{center}
		\includegraphics[width=1\textwidth]{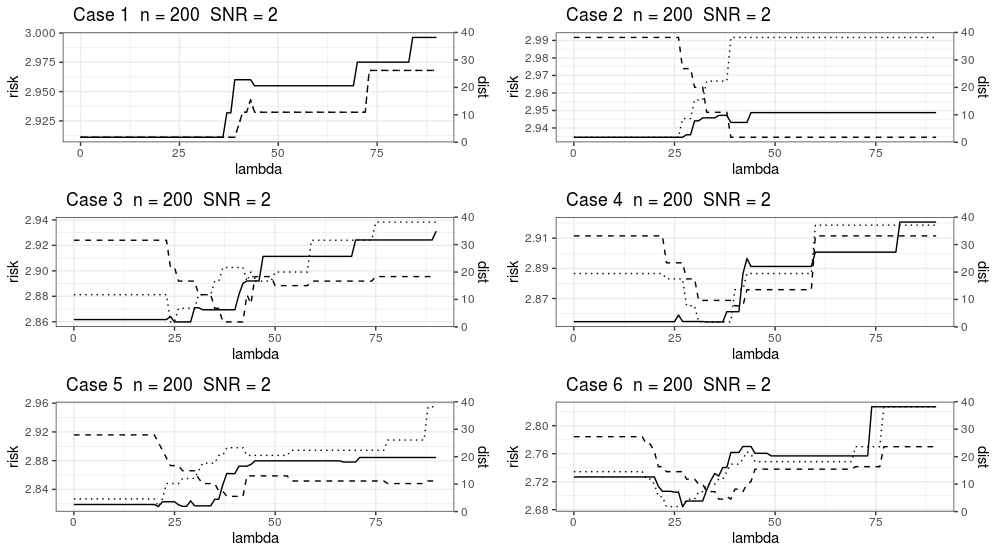}
	\end{center}
	\caption{The values of empirical risk (solid) and $d( \widehat{\mathfrak{S}}_{tr}(\widetilde{\lambda}), \widehat{\mathfrak{S}}(\lambda)$ (dotted), $d (\widehat{\mathfrak{S}} (\lambda), \mathfrak{S} )$, (dashed).
\label{Pen.em.2}}
\end{figure}

Consider next the signal-to-noise ratio is small, $\textrm{SNR} = 2$. Unfortunately, the empirical risk is minimized at smaller values of $\lambda$ than desired; see Fig.~\ref{Pen.em.2}. Unlike Fig. \ref{fig:second}, solid line (empirical risk) shows a far different shape than dashed line (the measure of dissimilarity), except for Model 1, which implies that empirical risks fail to detect the true structure. This is due to the lower value of SNR, with which the magnitude of noise overwhelms that of signal. As the score vectors of each dataset have almost random directions in low SNRs, there is a tendency that signals from a partially-joint (and fully-joint) scores are counted separately as if they belong to individual data blocks.

%%% n = 200, SNR = 2
%\begin{figure}[h]
%	\centering
%	\begin{tabular}{cc}
		%\subfloat{\includegraphics[width=.45\linewidth]{./image/m1.n200.SNR2}} &
		%\subfloat{\includegraphics[width=.45\linewidth]{./image/m2.n200.SNR2}} \\
		%\subfloat{\includegraphics[width=.45\linewidth]{./image/m3.n200.SNR2}} &
		%\subfloat{\includegraphics[width=.45\linewidth]{./image/m4.n200.SNR2}} \\
		%\subfloat{\includegraphics[width=.45\linewidth]{./image/m5.n200.SNR2}} &
		%\subfloat{\includegraphics[width=.45\linewidth]{./image/m6.n200.SNR2}} \\
	%\end{tabular}
	%\caption{The values of penalized empirical risk for $\lambda$ when $n = 200$ and $\textrm{SNR} = 2$.  The values of empirical risk (solid line) and the measure of dissimilarity, ${\rm diff}(\widehat{\mathfrak{S}}_0 (\lambda, \widehat{Z}), \mathfrak{S}_0 )$, (dashed line) over varying $\lambda$ are shown.}
	%\label{Pen.em.2}
%\end{figure}

\clearpage

\subsection{Simulation Settings for Section 5.4}

In the unbalance of signal strength between joint and individual component settings, we set $n = 200$ and $K = 3$.

In the first case, we set the elements of $\sigma_M^2$ as
\begin{itemize}
	\item[(1)] Joint ($S_1$) : $(15, 14.5, \ldots, 5.5)$,
	\item[(2)] Individual 1 ($S_2$) : $(0.150, 0.141, 0.132, \ldots, 0.069)$,
	\item[(3)] Individual 2 ($S_3$) : $(0.147, 0.138, 0.129, \ldots, 0.066)$,
	\item[(4)] Individual 3 ($S_4$) : $(0.144, 0.135, 0.126, \ldots, 0.063 )$,
\end{itemize}
so the strength of joint signals are about 100 times stronger than those of individual signals.

In the second case case, we set the elements of $\sigma_M^2$ as
\begin{itemize}
	\item[(1)] Joint ($S_1$) : $(0.15, 0.145, \ldots, 0.055)$,
	\item[(2)] Individual 1 ($S_2$) : $(15, 14.1, 13.2, \ldots, 6.9)$,
	\item[(3)] Individual 2 ($S_3$) : $(14.7, 13.8, 12.9, \ldots, 6.6)$,
	\item[(4)] Individual 3 ($S_4$) : $(14.4, 13.5, 12.6, \ldots, 6.3 )$,
\end{itemize}
so the strength of individual signals are about 100 times stronger that those of joint  signals.

\clearpage

\section{Additional Information on Real Data Analysis}

\setcounter{table}{0}

\begin{table}[h]
		\caption{Fisher's exact tests between gene mutations/chromosome defects and the CLL subgroups, $\alpha$ and $\beta$ were conducted simultaneously. The p-values were adjusted with the FDR (Benjamini-Hochberg method) and the top 5 lowest FDR (BH method) adjusted p-values were listed.}
	\label{RDF.table1}
	\centering
	\begin{tabular}{ |c|c|c|  }
		\hline
		Gene mutation/Chromosome defect &  adjusted p-value \\
		\hline
		\textbf{IGHV}   & $\mathbf{1.036 \times 10^{-13}}$ \\
		MED12  & $0.173$ \\
		del17p13  & $0.174$ \\
		del13q14  & $0.178$ \\
		TP53  & $0.184$ \\
		\hline
	\end{tabular}
\end{table}

\begin{table}[h]
		\caption{The association between IGHV mutation status and the CLL subgroups with an FDR (BH method) adjusted p-value $\mathbf{1.036 \times 10^{-13}}$ from the Fisher's exact test. Nine missing data for IGHV mutation status were excluded.}
	\label{RDF.table_2}
	\centering
	\begin{tabular}{ |c|c|c|  }
		\hline
		Mutation $\backslash$ Subgroup &  $\alpha$  & $\beta$ \\
		\hline
		IGHV Wild type  & 49 & 8 \\
		IGHV Mutated  & 7   & 48 \\
		\hline
	\end{tabular}
\end{table}

\begin{table}[bt]
	\caption{The most appeared drugs in the subgroups [a] and [b] with appearances at least four times out of 5 concentrations.}
	\label{tab:tabfive}
	\centering
	\scriptsize
	\begin{tabular}{ c|c|c  }
		\hline
		\multicolumn{3}{c}{The most appeared drugs in subgroup [a]} \\
		\hline
		Drug name &  Target pathway & Appearances \\
		\hline
		spebrutinib   &  BTK & 5 \\
		idelalisib  & PI3K delta & 5 \\
		duvelisib  & PI3K gamma, PI3K delta & 5 \\
		tamatinib & SYK & 5 \\
		dasatinib & ABL1, KIT, LYN, PDGFRA, PDGFRB, SRC    & 5  \\
		PF 477736    &  CHK1, CHK2   & 5 \\
		MK-2206    & AKT1/2 (PKB) & 5 \\
		\hline
		ibrutinib  & BTK & 4 \\
		selumetinib   & MEK1/2     & 4 \\
		PRT062607 HCL    & SYK & 4 \\
		AZD7762     & CHK1/2    & 4 \\
		CCT241533    & CHK2 & 4 \\
		TAE684   & ALK & 4 \\
		MK-1775      & WEE1 & 4 \\
		AT13387    & HSP90 & 4 \\
		\hline
	\end{tabular}
	
	\vspace{1.5cm}
	
	\begin{tabular}{ c|c|c  }
		\hline
		\multicolumn{3}{c}{The most appeared drugs in the subgroup [b]} \\
		\hline
		Drug name &  Target pathway & Appearances \\
		\hline
		everolimus   & mTOR & 5 \\
		thapsigargin  & SERCA & 5 \\
		orlistat  & LPL & 5 \\
		rotenone & Electron transport chain in mitochondria & 5 \\
		\hline
		afatinib & EGFR, ERBB2 & 4 \\
		fludarabine & Purine analogue & 4 \\
		navitoclax & BCL2, BCL-XL, BCL-W & 4 \\
		\hline
	\end{tabular}
\end{table}

\clearpage

\bibliographystylesm{elsarticle-harv}
\bibliographysm{Bibliography-MM-MC}

\end{document}